\def\draftversion{false}
\def\showall{true}  % Set to false to hide figures

\RequirePackage{ifthen}
\ifthenelse{\equal{\draftversion}{true}}{
  \documentclass[aps,pra,10pt,galley,amsmath,amssymb,
                 longbibliography,
                 superscriptaddress,nofootinbib]{revtex4}
}{
     \documentclass[aps,pra,10pt,twocolumn,amsmath,amssymb,floatfix,
    longbibliography,
    superscriptaddress,nofootinbib
    	]{revtex4-1}
}

\usepackage{amsmath,amssymb,amsfonts,mathtools}
\usepackage{amsthm}
\usepackage{nicefrac}
\usepackage{graphicx}
\usepackage{hhline}
\usepackage[utf8]{inputenc}
\usepackage{subfigure}
\usepackage{xcolor}
\usepackage{bbm}
\usepackage{ulem}
\usepackage{mathrsfs}
\usepackage{tabularx}
\usepackage{soul}  % \st    for strike-out
\usepackage{hyperref}
\hypersetup{
    colorlinks=true,       
    linkcolor=blue,          
    citecolor=blue,       
    filecolor=magenta,      
    urlcolor=blue          
}

\ifthenelse{\equal{\draftversion}{true}}{
  \marginparwidth 2.7in
  \marginparsep 0.5in
  \newcounter{comm} % counter for commentaries
  \def\commnext{\stepcounter{comm}}
  \def\commtext{{\bf\color{blue}[\arabic{comm}]}}
  \def\commmar{{\bf\color{blue}[\arabic{comm}]}}
  \def\stm#1{\commnext\marginpar{\small ST\commmar: #1}\commtext}
  \def\ism#1{\commnext\marginpar{\small IS\commmar: #1}\commtext}
   \def\xlm#1{\commnext\marginpar{\small XL\commmar: #1}\commtext}
  
  \def\parsedate #1:20#2#3#4#5#6#7#8\empty{20#2#3/#4#5/#6#7}
  \def\moddate{\expandafter\parsedate\pdffilemoddate{\jobname.tex}\empty}
  \newcommand{\mydate}{\date[Last modified ]{\moddate}}
  \def\tenxl#1{\textcolor{blue}{#1}}   % for tent changes by xl
\def\tents#1{\textcolor{magenta}{#1}}   % for tent changes by ST
}{
  \def\stm#1{}
  \def\ism#1{}
  \def\xlm#1{}

  \newcommand{\mydate}{\date{\today}}
  \def\tenxl#1{#1}   % for tent changes by xl
\def\tents#1{#1}   % for tent changes by ST
}

\ifthenelse{\equal{\showall}{false}}{
  \renewcommand\includegraphics[2][]{{\bf Figure not shown.}}
  \renewcommand\input[1][]{{\bf Not shown\ }}
}{}

\newcommand{\beq}{\begin{equation}}
\newcommand{\beqo}{\begin{equation} \setlength{\abovedisplayskip}{0pt} \setlength{\belowdisplayskip}{0pt}}
\newcommand{\eeq}{\end{equation}}
\newcommand{\bea}{\begin{eqnarray}}
\newcommand{\eea}{\end{eqnarray}}
\newcommand{\beas}[1]{\begin{subequations}\eql{#1}\begin{eqnarray}}
\newcommand{\eeas}{\end{eqnarray}\end{subequations}}
\newcommand{\nn}{\nonumber\\}

\newcommand{\eq}[1]{Eq.~(\ref{eq:#1})}
\newcommand{\Eq}[1]{Equation~(\ref{eq:#1})}
\newcommand{\eqs}[2]{Eqs.~(\ref{eq:#1}) and (\ref{eq:#2})}

\newcommand{\Eqs}[2]{Equations~(\ref{eq:#1}) and (\ref{eq:#2})}

\newcommand{\eqr}[2]{Eqs.~(\ref{eq:#1}-\ref{eq:#2})}
\newcommand{\Eqr}[2]{Equations~(\ref{eq:#1}-\ref{eq:#2})}

\newcommand{\fref}[1]{Fig.~\ref{fig:#1}}

\newcommand{\sref}[1]{Sec.~\ref{sec:#1}}

\newcommand{\aref}[1]{Appendix~\ref{sec:#1}}

\newcommand{\tref}[1]{Tab.~\ref{tab:#1}}

\ifthenelse{\equal{\draftversion}{true}}{
  \newcommand{\eql}[1]{\label{eq:#1}\hbox{\Red{\small\;\;[#1]}}}
  \newcommand{\figl}[1]{\label{fig:#1}\Red{\small\;\;[Fig:~#1]}}
  \newcommand{\secl}[1]{\label{sec:#1}\Red{\small\;\;[Sec:~#1]}}
  \newcommand{\tabl}[1]{\label{tab:#1}\Red{\small\;\;[Tab:~#1]}}
}{}
\ifthenelse{\equal{\draftversion}{false}}{
  \newcommand{\eql}[1]{\label{eq:#1}}
  \newcommand{\figl}[1]{\label{fig:#1}}
  \newcommand{\secl}[1]{\label{sec:#1}}
  \newcommand{\tabl}[1]{\label{tab:#1}}
}{}

\newcommand{\dis}{\displaystyle}

\newcommand{\ket}[1]{\vert#1\rangle}
\newcommand{\bra}[1]{\langle#1\vert}
\newcommand{\ip}[2]{\langle#1\vert#2\rangle}
\newcommand{\me}[3]{\langle#1\vert#2\vert#3\rangle}

\def\memnR#1{\me{\mathbf{0}i}{#1}{\RR j}}

\newcommand{\wt}[1]{\widetilde{#1}}

\newcommand{\Real}[1]{\rm Re\,}
\newcommand{\Imag}[1]{\rm Im\,}
\newcommand{\tr}[1]{\rm tr\,}
\newcommand{\Tr}[1]{\rm Tr\,}

\def\wt#1{\widetilde{#1}}
\def\Re{\mathrm{Re}}
\def\Im{\mathrm{Im}}
\def\Tr{\mathrm{Tr}}

\def\k{{\bf k}}
\def\m{{\bf m}}

\def\rr{{\bf r}} % (avoid \r, messes up with \AA=\r{A})
\def\R{{\bf R}}
\def\0{{\bf 0}}

\def\bt{{\bf t}}

\def\bOmega{{\boldsymbol\Omega}}
\def\bnabla{{\boldsymbol\nabla}}
\def\w{\ww}

\def\kk{{\bf k}}
\def\rr{{\bf r}}
\def\RR{{\bf R}}
\def\dk{[d\kk]}

\def\bnk{_{\k n}}
\def\bmk{_{\k m}}

\def\unk{u\bnk}
\def\enk{\varepsilon\bnk}
\def\emk{\varepsilon\bmk}

\def\en{\varepsilon_n}
\def\el{\varepsilon_l}
\def\ej{\varepsilon_j}

\def\abc{\varepsilon_{abc}}
\def\dk{[d\k]}

\def\la{\langle\kern-2.5pt\langle}
\def\ra{\rangle\kern-2.5pt\rangle}
\def\vt{\vert\kern-1.5pt\vert}

\def\lip#1#2{\la#1\vt#2\ra}
\def\lme#1#2#3{\la#1\vt#2\vt#3\ra}

\def\AB{{\rm AB}}

\def\all{^{\rm all}}

\def\one{{\mathbbm{1}}}

\def\ww{^{\text{W}}}

\def\abc{\epsilon_{abc}}
\def\conjabmn{\left(_{m\leftrightarrow n}^{a\leftrightarrow b}\right)^* }

\def\inn{^{A}}
\def\out{^{B}}
\def\all{}  % by default summation is over all

\def\Ham{{\cal H}}
\def\ef{\epsilon_{\rm F}}

\def\Fcov{\wt F}
\def\Fbar{\overline{F}}
\def\Fwan{\left( F\ww \right)}

\def\Hcov{\wt H}

\def\Hbar{\overline{H}}
\def\Hwan{\left( H\ww \right)}

\def\Abar{\overline{A}}

\def\eff{}
\newcommand{\Hcoma}[3]{\Ham_{#2#3}^{,#1}}
\newcommand{\Hcomatwo}[3]{\Hcoma{#1}{#2}{#3}}

\def\zerot#1{\breve{#1}}
\begin{document}

\title{
    Covariant derivatives of Berry-type % curvature and other geometric
    quantities: Application to nonlinear transport % in solids
  }
%
  % Gauge-covariant $k$ derivatives of Berry-phase quantities for
  % nonlinear conductivities
  %
  % Evaluation of intraband nonlinear conductivities by Wannier interpolation
  %
  % Covariant formulation of $k$-space derivatives of
  % Berry curvature: and related quantities
  %
  % Wannier-based
  % % {\it ab initio}
  % implementation of the Berry-Boltzmann formalism for nonlinear transport
  %
  % Wannier formulation of the Berry-Boltzmann formalism for nonlinear
  % transport

\author{Xiaoxiong Liu} \affiliation{Physik-Institut, Universit\"at
  Z\"urich, Winterthurerstrasse 190, CH-8057 Z\"urich, Switzerland}

\author{Stepan S. Tsirkin} \affiliation{Centro de F{\'i}sica de Materiales,
  Universidad del Pa{\'i}s Vasco, 20018 San Sebasti{\'a}n,
  Spain} \affiliation{Ikerbasque Foundation, 48013 Bilbao, Spain}

\author{Ivo Souza} \affiliation{Centro de F{\'i}sica de Materiales,
  Universidad del Pa{\'i}s Vasco, 20018 San Sebasti{\'a}n,
  Spain} \affiliation{Ikerbasque Foundation, 48013 Bilbao, Spain}

\begin{abstract}
The derivatives of the Berry curvature $\Omega$ and intrinsic orbital magnetic moment $\m$ in momentum space 
are relevant to various problems, including the nonlinear anomalous Hall effect and magneto-transport within the Boltzmann-equation formalism. 
To investigate these properties using first-principles methods, we have developed a Wannier interpolation scheme that evaluates the 
``covariant derivatives'' of the non-Abelian $\Omega$ and $\m$ matrices for a group of bands within a specific energy range of interest. 
Unlike the simple derivative, the covariant derivative does not involve couplings within the groups and preserves the gauge covariance of the $\Omega$ and $\m$ matrices. 
In the simulation of nonlinear anomalous Hall conductivity, the resulting ``Fermi-sea'' formula for the Berry curvature dipole are more robust and converges faster with the density of the integration k-grid than the ``Fermi-surface'' formula implemented earlier. The developed methodology is made available via the open-source code WannierBerri and we demonstrate the efficiency of this method through first-principles calculations on trigonal Tellurium.
\end{abstract}
\mydate
\maketitle

%======================
\section{Introduction}
\secl{intro}
%======================

\ism{Maybe start by saying that covariant derivatives in $\k$ space
  play a central role in the microscopic theory of nonlinear {\it
    optical} responses.  Here, we show that nonlinear {\it transport}
  responses are also most naturally formulated in terms of covariant
  derivatives of geometric quantities characterizing the occupied
  Bkloch states.}

The absence of spatial inversion and/or time-reversal symmetry gives
rise to various nonlinear
% charge-
transport phenomena in
solids~\cite{tokura-nc18},
\ism{Add the two 2021 reviews here?}
including
unidirectional magnetoresistance~\cite{rikken-prl01,avci-natphys15} and
nonlinear Hall
effects~\cite{deyo-arxiv09,sodemann-prl15,ma-nature19,kang-natmater19,huang-arxiv21,zhang-arxiv20}.
% \Red{Addititonal
%   nonlinear effects appear if time-reversal symmetry is also
%   broken. An example is unidirectional
%   magnetoresistance~\cite{avci-natphys15}, which can be viewed as a
%   ``spontaneous'' version of the magnetochiral anisotropy induced by a
%   magnetic field.}
%
Such effects are encoded in the expansion of the
% charge-
current density in powers of the applied electric and magnetic fields,
\begin{align}
j_a&=\sigma^{10}_{ab}E_b+\sigma^{11}_{ab\alpha}E_b B_\alpha+
\sigma^{12}_{ab\alpha\beta}E_b B_\alpha B_\beta \nn
&+\sigma^{20}_{abc}E_bE_c+\sigma^{21}_{abc\alpha}E_b E_c B_\alpha \nn
&+\sigma^{30}_{abcd}E_bE_cE_d \nn
&+\ldots\,.
\eql{sigma-expansion}
\end{align}
%
% where the subscripts denote Cartesian indices.
%
% In the limit of static or low frequency fields,
At low frequencies,
the expansion coefficients
% \Red{(magneto)}
% conductivity tensors
% % appearing
% in \eq{sigma-expansion} 
can be evaluated using
semiclassical~\cite{xiao-rmp10,gao-fp19} as well as fully
quantum-mechanical methods~\cite{morimoto-prb16,watanabe-prr20}. The
resulting expressions contain two types of terms:
%
% \ism{The terms ``intraband'' and ``interband'' usually mean something
%   different: intraband vs interband optical conductivity, depending on
%   how the optical frequency compares with the fundamental gap. Do we
%   need to clarify that this is {\it not} what we mean?}
%
(i)
intraband terms that only involve intrinsic geometric properties of
the unperturbed Bloch states (Berry curvature, effective mass,
magnetic moment, and quantum metric);
%
% \ism{This definition of intraband terms includes the ``horizontal
%   mixing'' corrections depicted in Fig.~1 of
%   Ref.~\cite{gao-fp19}. These contain the quantum metric as well as
%   its first $\k$ derivative, which appears in the Christoffel symbol.}
%
% and
(ii) interband terms that take into account how the Bloch states
change, via vertical band mixing, under the applied
fields~\cite{gao-fp19}.

This paper deals with the evaluation of intraband nonlinear
conductivities, which typically contain {\it derivatives} of
% Berry curvature
% (and related quantities)
geometric quantities with respect to crystal momentum $\k$.  While the
focus will be on {\it ab initio} implementations based on Wannier
functions, the formalism presented here can also be combined with
effective-Hamiltonian methods such as $\k\cdot{\bf p}$ and
tight-binding.

% As a concrete example,
Consider the current response at zero magnetic field,
\beq
j_a=\sigma^{10}_{ab}E_b+\sigma^{20}_{abc}E_bE_c+
\sigma^{30}_{abcd}E_bE_cE_d+\ldots
\eql{sigma-expansion-E}
\eeq
Working in the constant relaxation time approximation and neglecting
interband contributions, one obtains~\cite{zhang-arxiv20}
%
% \ism{By writing the Berry curvature as an antisymmetric tensor these
%   expressions acquire a nice symmetry, I think. But later on we can
%   convert to axial-vector form, if it simplifies things. (The vector
%   form is useful to discuss the tracelessness of the Berry dipole, for
%   example.)}
%
\begin{align}
\sigma^{10}_{ab}&=\frac{e^2}{\hbar}\int\dk\sum_n\,f_0(\enk)
\left[(\tau/\hbar)\partial^2_{ab}\enk -\Omega^{ab}\bnk\right]\,,
\eql{sigma-10}\\
\sigma^{20}_{abc}&=\frac{e^3\tau}{\hbar^2}\int\dk\sum_n\,f_0(\enk)
\left[-(\tau/\hbar)\partial^3_{abc}\enk+\partial_c\Omega^{ab}\bnk\right]\,,
\eql{sigma-20}\\
\sigma^{30}_{abcd}&=\frac{e^4\tau^2}{\hbar^3}\int\dk\sum_n\,f_0(\enk)
\bigg((\tau/\hbar)\partial^4_{abcd}\enk\nn
&\,\,\,\,\,\,\,\,\,\,\,\,\,\,\,\,\,\,\,\,\,\,\,\,\,\,\,\,\,\,\,\,\,\,\,\,\,\,
\,\,\,\,\,\,\,\,\,\,
%\,\,\,\,\,\,\,\,\,
-\frac{3}{4}\partial^2_{cd}\Omega^{ab}\bnk
-\frac{1}{4}\partial^2_{bc}\Omega^{ad}\bnk
\bigg)\,.
\eql{sigma-30}
\end{align}
where $\tau$ is the relaxation time, $\dk\equiv d^3k/(2\pi)^3$,
$\partial_a\equiv\partial/\partial k_a$, and the integrals are over
the first Brillouin zone (BZ). The band energy is denoted $\enk$, $f_0(\enk)$
is the Fermi-Dirac distribution function,
and
\beq
\Omega^{ab}\bnk=-2\Im\ip{\partial_a\unk}{\partial_b\unk}
% =-\calF^{ba}\bnk
\eql{curv-n-def}
\eeq
is the Berry curvature, where $\ket{\unk}$ is the periodic part of a
Bloch state $\ket{\psi\bnk}$. \tents{\Eqr{sigma-10}{sigma-30} are written in the so-called "Fermi-sea" form,
meaning that all states below the Fermi level contribute to the integral. }
%
%
% In writing
% these expressions, we have used the notation
% %
% \beq
% {\cal O}^{,a}\equiv\frac{\partial{\cal O}}{\partial k_a}\,,\quad
% {\cal O}^{,ab}\equiv\frac{\partial^2{\cal O}}{\partial k_a\partial k_b}\,,\quad
% {\cal O}^{ab,c}\equiv\frac{\partial{\cal O}^{ab}}{\partial k_c}
% \eql{der-notation}
% \eeq
% %
% for the partial derivatives of various
% % $\k$-dependent
% objects.

\Eq{sigma-10} gives the linear conductivity: the first term is the
Ohmic Drude conductivity expressed in terms of the inverse effective
mass of the occupied states; the second term, given by the net Berry
curvature of the occupied states, describes an intrinsic anomalous
Hall effect in magnetic conductors~\cite{nagaosa-rmp10}.
\Eq{sigma-20} gives the quadratic conductivity: the first term is
Ohmic and it describes unidirectional magnetoresistance in magnetic
acentric conductors~\cite{zelezny-arxiv21}, while the second term
describes an anomalous Hall effect in nonmagnetic acentric
conductors~\cite{deyo-arxiv09,sodemann-prl15,ma-nature19,kang-natmater19}.
\Eq{sigma-30} describes cubic Ohmic and anomalous Hall responses,
which so far have only been studied
theoretically~\cite{parker-prb19,zhang-arxiv20}.  Note that
higher-order conductivities contain higher-order derivatives
of % either
the band dispersion
% or
and of the Berry
curvature.
% $\enk$ and $\Omega^{ab}\bnk$.
For magnetoconductivities such as
$\sigma^{11}_{ab\alpha}$, $\sigma^{12}_{ab\alpha\beta}$ and
$\sigma^{21}_{abc\alpha}$ in \eq{sigma-expansion}, the intrinsic
magnetic moment
% of the Bloch states
% ${\bf m}\bnk$
and its derivatives are needed as
well~\cite{xiao-rmp10,gao-fp19,morimoto-prb16,lahiri-arxiv21}.
\ism{Added citation to the recent preprint that worked out the $E^2B$
  Berry-Boltzmann terms.}
%
% \ism{I guess the quantum metric will appear as well at higher orders
%   in the fields, do you agree?}

% First-principles calculations of nonlinear conductivities and
% magnetoconductivites are still in their infancy, calling for the
% development of improved methodologies.
The first-principles evaluation of both terms in the linear
conductivity~\eqref{eq:sigma-10} is by now a fairly routine task.
% In the case of the linear conductivity given by \eq{sigma-10}, it is
% well known how to compute the needed ingredients from first
% principles.
A popular approach
% that is both accurate and efficient
is
Wannier interpolation~\cite{wang-prb06,yates-prb07}, where a
Slater-Koster type of interpolation is carried out for the quantities
of interest after mapping the low-energy {\it ab initio} electronic
structure onto a basis of localized Wannier functions.

When it comes to nonlinear conductivities, {\it ab initio}
calculations are still quite recent.
In the case of intraband
responses such as \eqs{sigma-20}{sigma-30}, a possible strategy is as
follows. First compute the inverse effective mass and Berry curvature
on a dense $\k$ mesh by Wannier interpolation, and then evaluate their
$\k$ derivatives by finite differences. This strategy was used in
several recent studies of nonlinear anomalous Hall
effects~\cite{zhang-prb18,zhang-2dmater18,zhang-arxiv20,he-arxiv21}.

\tents{Another common strategy is to employ integration by parts in 
\eqr{sigma-20}{sigma-30} and thus transfer the derivative  
from the Berry curvature $\Omega^{ab}\bnk$ to the Fermi-Dirac distribution
$f_0(\enk)$. For instance, the second term of \eq{sigma-20} is governed by 
the so-called "Berry curvature dipole"~\cite{sodemann-prl15} which in the Fermi-sea formulation is 
given by
\beq
\mathcal{D}_{cd}^{\rm sea} = \epsilon_{abd} \int \dk \sum_n \partial_c \Omega_{\k n}^{ab} f_0(\enk)~,
\eql{dp-sea}
\eeq 
Using integration by parts it may be rewritten in as 
\beq
\mathcal{D}_{cd}^{\rm surf} = \epsilon_{abd} \int \dk \sum_n  \Omega_{\k n}^{ab} \partial_c\enk  \left(- \frac{\partial f_0}{\partial \varepsilon}\right)_{\varepsilon = \enk}  ~,
\eql{dp-surf}
\eeq 
At low temperature the derivative of the distribution function $f_0'$ is a narrow peak, which ensures that only electronic states that are close to the Fermi level contribute to the integral. Therefore, such formulations are called ``Fermi-surface'' integrals. 
As we will show, in numerical simulations, such integrals require denser sampling of the Brillouin zone, compared to the ``Fermi-sea'' integrals. Moreover, for magnetoconductivities magnetoconductivities such as
$\sigma^{11}_{ab\alpha}$, $\sigma^{12}_{ab\alpha\beta}$ and
$\sigma^{21}_{abc\alpha}$ it is not always possible to get rid of derivatives of Berry curvature and orbital magnetic moment simultaneously. 
}

In this work, we develop an alternative approach where
derivatives of geometric quantities
% such
% as $\partial_c\calF^{ab}$
are evaluated perturbatively,
% at each grid point,
% by Wannier interpolation,
without resorting to finite differences.  Importantly, the expressions
we obtain are oblivious to band crossings and avoided crossings away
from the Fermi level, as expected on physical grounds.  We emphasize
that this is not the case if one differentiates 
% for the expressions obtained via a
% straightforward differention of
% the formulas for
the Berry curvature in a naive way:
% and related geometric quantities:
the resulting expression contains spurious terms that react strongly
to remote level crossings (see for example
Ref.~\cite{morimoto-prb16}).  To obtain well-behaved gradient formulas
we start from gauge-covariant matrix objects such as the non-Abelian
Berry curvature, and differentiate them using a 
%generalized derivative
``covariant derivative''
\ism{I think we should just call it ``covariant derivative'' (see my
  notes from November 2021).}
that preserves gauge covariance.  The nonlinear conductivities are
then expressed as gauge-invariant traces.
% of the resulting gauge-covariant matrices.

The manuscript is structured as follows: \tents{In \sref{preliminary}, 
we discuss the calculation of the derivative of the Berry curvature 
for an effective model in order to demonstrate the essence of the problem. 
In \sref{generalized-der} we introduce the covariant derivative and discuss its properties. 
In \sref{geometric-quantity} we demonstrate, how multiple geometrical quantities of interest may be formulated in a gauge-covariant way and reduced to two objects $\Fcov^{ab}_{mn}$ and  $\Hcov^{ab}_{mn}$. 
\sref{wannier-interpolation} explains how to evaluate those objects using Wannier interpolation, or for a tight-binding model. 
Finally, in \sref{result-Te}, we present the first-principle simulation of the Berry curvature dipole in Te, where we show that the "Fermi-sea" formula with the covariant gradient of the Berry curvature has better convergence and is more robust than the "Fermi-surface" formula.
}
%%% Local Variables:
%%% mode: latex
%%% TeX-master: "pap"
%%% End:

%=================================
\section{Statement of the problem}
\secl{preliminary}
%=================================
%
% \ism{Previous title: ``Preliminary discussion''. Added two
%   subsections. OK?}

\begin{figure}
\begin{center}
\includegraphics[width=0.95\columnwidth]{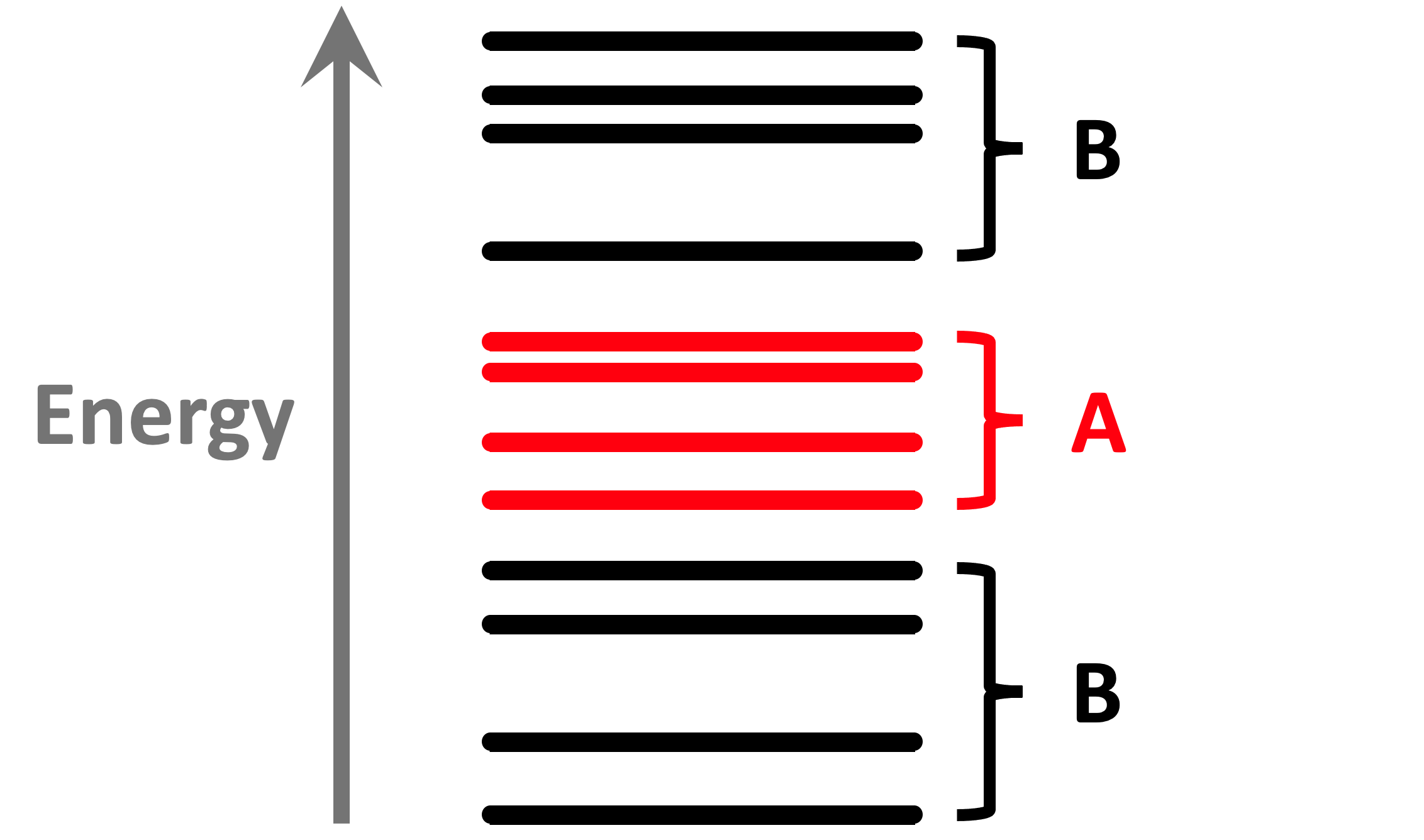}
\caption{Partition of the energy levels of an effective Hamiltonian
  $\Ham_\k$ into an ``active'' group A containing the levels of
  interest, and its complement B. The two groups are separated in
  energy, but degeneracies may be present within each group.}
\figl{A-B}
\end{center}
\end{figure}

Before detailed derivations, let us present our scheme in the
simplest possible setting.  We consider a system described by a
% $\k$-dependent
an effective Hamiltonian $\Ham\eff_\k$ (e.g., a tight-binding or
$\k\cdot\mathbf{p}$ Hamiltonian) with a finite number of
eigenvectors $\ket{m\k}$ and eigenvalues $\emk$ at each $\k$,
\ism{Later on, when specializing to TB models and Wannier
  interpolation, the effective Hamiltonian $\Ham\eff_\k$ will be
  denoted as $H\w_\k$. Here I want to distinguish it from the {\it ab
    initio} Bloch Hamiltonian $\Ham_\k$. or maybe we should denote the
  latter as $\Ham_\k^{\rm KS}$ (Kohn-Sham) and its eigenvalues
  $\enk^{\rm KS}$, for extra clarity.}
\stm{I do not see a need to distinguish between the "effective" and "Kohn Sham" Hamiltonians. I suggest to remove "eff" supoerscript here also}

\beq
\Ham\eff_\k\ket{m\k} = \emk\ket{m\k}\,.
\eeq
In practice, one is typically interested in groups of eigenstates, not
in individual states. Let us denote the ``active'' group of
interest by A and its complement by B, as depicted in \fref{A-B}. It
is assumed that the A and B groups are well separated in energy, but
degeneracies may be present within each of them. The completeness relation is 
\beq
\one=\sum_m^{\rm A}\,\ket{m}\bra{m}+\sum_{l}^{\rm B}\,\ket{l}\bra{l} \equiv\sum_j^{\rm all}\,\ket{j}\bra{j} \,.
\eql{completeness}
\eeq
In this paper we will consistently use indices $i,j,j',\ldots$ to 
denote the whole set of Wannier states, while $m,n,n',n'',\ldots$ will denote states 
of subspace A, and $p,l,l',l'',\ldots$ for states in subspace B.
We will assume that the \textit{repeated primed}
indices are summed (with the values running over the corresponding subspace), while the non-primed
indices are not summed unless written explicitly (for the Trace quantities).
Thus, from \sref{geometric-quantity}, we shorten the equation by omiting the $\sum$ symbols.

% (we will refer to them as ``composite groups'').
% For \eqr{sigma-10}{sigma-30}, groups A and B at a given $\k$ comprise
% all states below and above the Fermi level, respectively.

%%% Local Variables:
%%% mode: latex
%%% TeX-master: "pap"
%%% End:

% As a first example,
Let us start with the linear anomalous Hall conductivity given by
second term in \eq{sigma-10}. To evaluate it we need the net Berry
curvature $\Omega^{ab}_{\rm A}\equiv\sum_n^{\rm A}\,\Omega^{ab}$ of the
occupied (A) states at each $\k$, where the Berry curvature of an
individual state is
$\Omega^{ab}_n=-2\Im\lip{\partial_a n}{\partial_b n}$ according to
\eq{curv-n-def} (the index $\k$ has been omitted for brevity). Using
the standard result from first-order non-degenerate perturbation theory, one
obtains~\cite{xiao-rmp10,vanderbilt-book18}
\beq
\ket{\partial_a n}=-i\alpha_n\ket{n}+
\sum_{j\not= n}\,\ket{j}
\frac{\me{j}{\partial_a \Ham\eff}{n}}{\en-\varepsilon_j}
\eql{dn-1}
\eeq
where $\alpha_n$ can be any real number.
%It is easy to find that the $l=n$ term does not contribute to 
%Berry Curvature. 
Inserting the completeness relation $\one=\sum_l\all\,\ket{j}\bra{j}$
The Berry curvature of a single band is 
\beq
\Omega_n^{ab}=-2\Im\sum_{j\not= n}\,
\frac{\me{n}{\partial_a \Ham\eff}{j}\me{j}{\partial_b \Ham\eff}{n}}
{\left(\en-\ej\right)^2}\,.
\eql{Omega-kn}
\eeq
%The summation contains $N-1$ terms, which is a small number for simple
%models. \Eq{Omega-kn} is therefore a very convenient means of evaluating the Berry curvature in
%effective models.

In typical applications one is more interested in the overall
properties of groups of energy eigenstates than in the properties of
individual eigenstates. For \eqr{sigma-10}{sigma-30}, groups A and B comprise all states
below and above the Fermi level at a given $\k$, respectively.
For the linear anomalous Hall conductivity 
%given by second term in \eq{sigma-10}; there, the quantity of interest is 
the net Berry curvature
%
%\beq
%\Omega^{ab}_{\rm A}=\sum_n^{\rm A}\,\Omega^{ab}_n
%\eeq
%
of the occupied states
%Using \eq{curv-n-def} and inserting \eq{completeness}
reads
%the double summation over states belonging to the same subspace (A or B) vanishes leaving
%
\beq
\Omega_{\rm A}^{ab}=-2\Im\sum_n^{\rm A}\sum_l^{\rm B}\,
\frac{\me{n}{\partial_a \Ham\eff}{l}\me{l}{\partial_b \Ham\eff}{n}}
{\left(\en-\el\right)^2}\,.
\eql{Omega-A-1}
\eeq
where the double summation over states belonging to set A cancels out.
Naturally, \eq{Omega-A-1} reduces to \eq{Omega-kn} when group A contains a single state.

By virtue of having $(\en-\el)^2$ in the denominator, \eq{Omega-A-1}
becomes resonantly enhanced in regions of the BZ where the gap between
strongly-coupled A and B states is small.  Conversely, the lack of
energy denominators involving pairs of A states or pairs of B states
means that $\Omega_{\rm A}^{ab}=-\Omega_{\rm B}^{ab}$ does not react
strongly to band crossings and avoided crossings within each sector.

All of the above are familiar results.
% properties of the Berry curvature.
Consider now the gradient of the Berry curvature,
% $\partial_c\calF^{ab}_{\rm A}$,
which
governs the quadratic anomalous Hall conductivity given by the second
term in \eq{sigma-20}. Like $\Omega_{\rm A}^{ab}$ itself,
$\partial_c\Omega_{\rm A}^{ab}$ should only react strongly to small
energy gaps between the two groups A and B, not
% to % internal
% (quasi)degeneracies
within each group. However, direct differentiation of \eq{Omega-A-1}
leads to an expression
% for $\partial_c\calF_{\rm A}^{ab}$
containing
energy denominators between pairs of states within the same group. The
problematic terms appear when differentiating the eigenstates
% in \eq{Omega-A-1}
using \eq{dn-1}.
% nondegenerate perturbation theory, which
% gives~\cite{vanderbilt-book18}
% %
% \ism{Replaced $n^\prime\rightarrow j$.}
% %
% \beq
% \lket{n_{,a}}=-i\alpha_n\lket{n}+
% \sum_{j\not= n}\,\lket{j}
% \frac{\lme{j}{H_{,a}}{n}}{\en-\epsilon_j}
% \eql{dn-1}
% \eeq
% %
% where $\alpha_n$ can be any real number.
If $n\in{\rm A}$, the summation over $j$ includes not only the B states
but also the other A states, which can be arbitrarily close in energy
to state $n$. Such unwanted terms cancel each other out in the final
result for $\partial_c\Omega_{\rm A}^{ab}$, but making that cancelation
explicit is not as straightforward as in the case of
\eq{Omega-A-1}. For $\partial^2_{cd}\Omega_{\rm A}^{ab}$ and higher
derivatives, achieving that cancellation becomes increasingly more
difficult.

To circumvent the above problem, we will develop a systematic
procedure for differentiating geometric quantities in such a way that
the spurious terms are absent by construction.
% In the case of
% When used
% to evaluate
% \eq{Omega-A-grad-def}
% $\partial_c\calF_{\rm A}^{ab}$,
% starting from \eq{Omega-A-1} for
% $\Omega_{\rm A}^{ab}$,
That procedure yields
\ism{Refer here to a later section where this expression is derived.}
%
%\begin{widetext}
%\begin{multline}
%\partial_c\Omega_{\rm A}^{ab}=-2\Im\sum_n^{\rm A}\sum_l^{\rm B}\,
%\frac{1}{\left(\en-\el\right)^2}
%\Bigg\{
%\me{n}{\partial_a \Ham\eff}{l}\me{l}{\partial^2_{bc}\Ham\eff}{n}\\
%+\me{n}{\partial_a\Ham\eff}{l}
%\Bigg[
%\sum_{n'}^{\rm A}\,
%\frac{\me{l}{\partial_b \Ham\eff}{n'}}{\el-\varepsilon_{n'}}\me{n'}{\partial_c \Ham\eff}{n}
%-\sum_{l'}^{\rm B}\,\me{l}{\partial_b \Ham\eff}{l'}
%\frac{\me{l'}{\partial_c \Ham\eff}{n}}{\varepsilon_{l'}-\en}
%+(b\leftrightarrow c)\Bigg]-(a\leftrightarrow b)\Bigg\}\,.
%\eql{Omega-A-grad-1}
%\end{multline}
%\end{widetext}
%
\tents{
\begin{widetext}
\begin{multline}
\partial_c\Omega_{\rm A}^{ab}=-2\Im\sum_n^{\rm A}\sum_l^{\rm B}\,
\frac{1}{\left(\en-\el\right)^2}
\Bigg\{
\Hcoma{a}{n}{l}\Hcomatwo{bc}{l}{n}
+\Hcoma{a}{n}{l}
\Bigg[
\sum_{n'}^{\rm A}\,
\frac{\Hcoma{b}{l}{n'} \Hcoma{c}{n'}{n}}{\el-\varepsilon_{n'}}
-\sum_{l'}^{\rm B}\,
\frac{\Hcoma{b}{l}{l'} \Hcoma{c}{l'}{n}}{\varepsilon_{l'}-\en}
+(b\leftrightarrow c)\Bigg]-(a\leftrightarrow b)\Bigg\}\,.
\eql{Omega-A-grad-1}
\end{multline}
\end{widetext}
%
% This sum-over-states formula
for the gradient of the Berry curvature
of an effective-Hamiltonian model, where we have used a shortened notation
$\Hcoma{a}{n}{l}\equiv\me{n}{\partial_a\Ham}{l}$
}
% is our first result.
Like \eq{Omega-A-1}, \eq{Omega-A-grad-1} contains energy denominators
between A and B states only, making it well-suited for numerical work.
\tents{Interestingly, \Eq{Omega-A-grad-1} is symmetric under $b\leftrightarrow c$, 
which is connected to the known property of the berry curvature dipole \eqs{dp-sea}{dp-surf}
to have zero trace. (see \aref{traceless} for details)}

\tents{In the following sections we will demonstrate how to systematically derive 
the derivatives of any order of Berry curvature, orbital magnetic moment,
quantum metric and similar quantities, and at the end of \sref{der-F-H} the derivation of \eq{Omega-A-grad-1} will emerge. }

%%% Local Variables:
%%% mode: latex
%%% TeX-master: "pap"
%%% End:

%\section{Non-Abelian generalized derivative}
\section{Non-Abelian covariant derivative}
\secl{generalized-der}
\ism{A key reference for this section is Ref.~\cite{mead-rmp92}, where
  the non-Abelian covariant derivative is discussed in
  Sec.~III.D. Note also that in his discussion of the non-Abelian
  formulation in Sec.~III, a central role is played by the operators
  $\hat{g}_\mu$ defined in Eq.~(3.15), whose matrix elements are
  nothing but the off-diagonal block of our $D^a$ matrix: see his
  Eq.~(4.22). Note also that Eq.~(3.18) gives an expression for
  $\partial_\mu \hat{g}_\nu$; we should try to relate it to our
  expression for $D^{a;b}_\AB$. See also below Eq.~(8) in
  arXiv:2110.13415, where they related it to the covariant derivative
  of the Bloch states (TO DO: work it out).}

\begin{table}[t]
\begin{center}
\caption{Covariant derivative and its useful properties. 
$m,n,n',n'',\ldots\in A$, $p,l,l',l'',\ldots\in B$, summation over repeated primed indices is implied.
See \sref{generalized-der} for details. }
       \begin{tabularx}{0.95\columnwidth}{ m{8cm}<{\centering} }
\hline
\hline
\textbf{Definition}\\
\hline
\beqo
    X^{:d}_{nl} \equiv \partial_d  X_{nl}+\,D^d_{nn'}X_{n'l}-\,X_{nl'}D^d_{l'l}
\eql{gender}
\eeq\\
\hline
\textbf{Matrix element of an operator $\hat{X}$}\\
\hline
\begin{subequations}
\bea
X_{nl}^{:d} &=& X^{,d}_{nl}  -
D^d_{nl'}X_{l'l}+ X_{nn'}D^d_{n'l} \,; \eql{gender-Xnl}\\
X_{ln}^{:d} &=& X^{,d}_{ln}  -
D^d_{ln'}X_{n'n}+ X_{ll'}D^d_{l'n} \,; \eql{gender-Xln}\\
X_{nm}^{:d}&=&X^{,d}_{nm}- D^d_{nl'}X_{l'm}+ X_{nl'}D^d_{l'm}\eql{gender-Xnn}\,,\\
X^{,d}_{nl}&\equiv& \me{n}{\partial_d\hat{X}}{l} \eql{operator-coma-derivative}
\eea\end{subequations}\\
\hline
\textbf{Hamiltonian $\Ham$}\\
\hline
\beqo
\Ham_{nm}^{:d} =\Ham_{nm}^{,d}\,, \quad \Ham_{nl}^{:d} = 0  \eql{H-gender}
\eeq\\
\hline
\hline
\textbf{Trace Rule}\\
\hline
\beqo
\partial_d\left( X_{n'n'}\right)= X_{n'n'}^{:d}
\eql{der-trace}
\eeq\\ 
\hline
\textbf{Product Rule}\\ 
\hline
        \beqo
        \left( X_{rs'}Y_{s't}\right)^{:d}=
        X_{rs'}^{:d}Y_{s't} + X_{rs'}Y_{s't}^{:d}  \eql{gender-prod-rule} \eeq\\
\hline
\textbf{Chain Rule}\\
\hline
\beqo
    \wt f_{ij}^{:d} = \Ham^{:d}_{ij}\times \Biggl\{
        \begin{array}{lll}
             \left. \frac{df(\epsilon)}{d\epsilon}\right|_{\epsilon=\epsilon_i}  & \mathrm{if}& \epsilon_i = \epsilon_{j}\\
             \frac{f(\epsilon_i)-f(\epsilon_{j})}{\epsilon_i-\epsilon_j} & \mathrm{if}& \epsilon_i\neq \epsilon_{j}
        \end{array}
\eql{gender-chain-rule}
\eeq\\
\hline\hline
\end{tabularx}

\tabl{property-cov-der}
\end{center}
\end{table}

%In this section, we will introduce covarant derivatives. In order to
%distinguish different derivatives more clearly, we use $\partial_a$ or
%subscript `$,a$' denotes simple derivatives, $\tilde{\partial_a}$,
%subscripts `$;a$' and `$:a$' denotes covarant derivatives.

%{\color{magenta} \sout{
%We will assume, that the \textit{repeated primed}
%indices are summed (with the values running over the corresponding subspace), while the non-primed 
%indices are not summed, unless written explicitly (for the Trace quantities). 
%Thus we shorten the equation by omiting the $\sum$ symbols.}
%}

%Before going ahead and differentiating \eq{Omega-tr-a} with respect to
%$\kk$, we need to introduce some more definitions and relations.
Consider two isolated groups of bands $A$ and $B$, and a matrix $X_{nl}$
% Eventually we will choose
% $A=\text{in}$ and $B=\text{out}$, but for now let us keep the
% discussion as general as possible.
The only assumption is that
$X_{nl}$ changes covariantly under gauge transformations $U_A$ and
$U_B$ that act separately on the $A$ and $B$ band groups,
\beq
\ket{n}\overset{U^A}{\longrightarrow}
\sum_m^A\,\ket{m}U^A_{mn}\,,\qquad
\ket{l}\overset{U^B}{\longrightarrow}
\sum_{p}^B\,\ket{p}U^B_{pl}\,.
\eql{gauge-transf}
\eeq
That is, we assume that
\beq
X_{nl}\overset{U^A}{\longrightarrow} 
\sum_m^A\,\left(U^A\right)^\dagger_{nm}X_{ml}\,,\qquad
X_{nl}\overset{U^B}{\longrightarrow} 
\sum_p^B\,X_{mp}U^B_{pl}\,.
\eeq
The problem is that the simple derivative $X^{,d}_{nl}$ is not
covariant in the above sense. This can be fixed by defining a
“covariant derivative” as in  \eq{gender} (see \tref{property-cov-der})
%
%\beq
%\begin{aligned}
%  %X^{:b}_{nl}&\equiv\partial_b X_{nl}+\sum_{n'}^A\,D^b_{nn'}X_{n'l}-
%  %\sum_{l'}^B\,X_{nl'}D^b_{l'l}
%    X^{:d}_{nl}&\equiv \partial_d  X_{nl}+\sum_{n'}^A\,D^d_{nn'}X_{n'l}-
%  \sum_{l'}^B\,X_{nl'}D^d_{l'l}
%\end{aligned}
%\eql{gender}
%\eeq
%
 % where the second definition is obtained from the first by
  % exchanging $n,n'\leftrightarrow l,l'$ and $A\leftrightarrow B$.
Note that if $X$ is Hermitian or anti-Hermitian, then $X^{:d}$ has the
same property. It can also be checked that
if $X$ is covariant then $X^{:d}$ is also covariant.

If group $A$ contains a single band $n$ and group $B$ a single band
$l$, we recover the Abelian definition of the covarant derivative
given in Eq~(9) of Ref.~\cite{aversa-prb95},
\beq
%X^{;b}_{nl}=\partial_b X_{nl}-i\left(A^b_{nn}-A^b_{ll}\right)X_{nl}\,,
X^{;d}_{nl}= \partial_d X_{nl} -i\left(A^d_{nn}-A^d_{ll}\right)X_{nl}\,,
\eql{gender-abelian}
\eeq
where we have written $D^a$ as $-iA^a$ for comparison purposes. To
avoid confusion, we denote the non-Abelian covarant derivative by
$X^{:d}$ and the Abelian one by $X^{;d}$.

Now let $\hat{X}$ be some operator. Then
\bea
X_{nl}^{:d}&\equiv & \me{n}{\hat{X}}{l}^{:d} \nonumber\\ &=&\me{n}{\partial_d\hat{X}}{l} + \me{\partial_d n}{\hat{X}}{l}+ \me{n}{\hat{X}}{\partial_d l} \nonumber\\
& &+\sum_{n''}^{A}D^d_{nn''}X_{n''l} - \sum_{l''}^{B} X_{nl''}D^d_{l''l} ,
\eea
and inserting the a completeness relation \eqref{eq:completeness} we get \eq{gender-Xnl}
Hereinafter  we will use a shortened notation \eqref{eq:operator-coma-derivative}
%\beq
%X^{,d}_{nl}\equiv \me{n}{\partial_d\hat{X}}{l} \eql{operator-coma-derivative}
%\eeq
%
If $n,l\in{\rm all}$ then only the first term survives, and our 
definition of the generalized derivative reduces to Eq.~(34) of Ref.~\cite{ventura-prb17}.
\eq{gender-Xnl} shows off-diagonal ``AB'' blocks of the matrix.
The diagonal block (``AA'') can be derived in a similar way to get \eq{gender-Xnn}
%
%\beq
%X_{nm}^{:d}=X^{,d}_{nm}-\sum_{l'}\out D^d_{nl'}X_{l'm}+\sum_{l'}\out X_{nl'}D^d_{l'm}\eql{gender-Xnn}\,,
%\eeq
%
and the ``BA'' and ``BB'' parts may be obtained by interchanging A and B in equations above.
One special case is the Hamiltonian operator $\hat \Ham$, which is represented by a diagonal matrix. Using $\Ham_{nl}=0$ and
$\Ham^{,d}_{nl}=D^d_{nl}(\varepsilon_l-\varepsilon_n)$ we
find the simple results given by \eq{H-gender}. 
% An obvious
% consequence of the equations above is
%
%\begin{subequations}
%\bea
%\Ham_{nm}^{:d}&=&\Ham_{nm}^{,d}\,,
%\eql{H-gender-in}\\
%\Ham_{nl}^{:d}&=& 0 
%\eea
%\end{subequations}
\tents{Thus we see, that although Hamiltonian is a diagonal matrix, its covariant derivative is only block-diagonal.}

It is important to show some useful properties of the covariant derivative, summarized in \tref{property-cov-der}.
\stm{do we really need a table, if we highlight the rules by subsection titles?}

\textbf{Trace rule.} First of all, in applications we will be interested in taking derivatives of a trace of matrix over subspace A. 
Using \eq{gender}, that can be written as 
\begin{multline}
\partial_d \sum_n X_{nn} = \sum_n \partial_d X_{nn} = \\
\sum_n  X^{:d}_{nn} + \sum_{nn'}^A\,D^d_{nn'}X_{n'n}-
  \sum_{nn'}^A\,X_{nn'}D^d_{n'n}
\end{multline}
and noting that the last to term are the same, upto the sign and interchange of indices $n\leftrightarrow n'$, we arrive at \eq{der-trace}

\textbf{Product rule.} 
Consider two covariant matrices $X_{rs}$ and $Y_{st}$ 
matrices with $r\in C_1$, $s\in C_2$ and $t\in C_3$ (where each subspace $C_i$ 
may independently  be equal to A or B). 
Taking the covariant derivative of their product explicitly we get
\begin{multline}
\left(\sum_s^{C_2} X_{rs}Y_{st}\right)^{:d}=\sum_s^{C_2}
\Bigl(  \left(\partial_d X_{rs}\right) Y_{st} +  X_{rs} \partial_d \left(Y_{st}\right)\\
+\sum_{r'}^{C_1} D^d_{rr'}X_{r's} Y_{st}-\sum_{t'}^{C_3} X_{rs} Y_{st'}D_{t't}^d  \Bigr)
\end{multline}
Now, adding and subtracting $\sum_{ss'}^{C_2} X_{rs} D_{ss'}^d  Y_{s't}$ we arrive at \eq{gender-prod-rule}, 
which is similar to the product rule for the simple derivatives.

\textbf{Chain rule.} 
In some cases equations may contain a scalar-valued smooth function of the electron energies $f(\epsilon_i)$. 
Taking the covariant derivatives of such functions is less intuitive than the product and trace rule. 
To do it correctly, we represent $f$ as a gauge-covariant matrix, 
\begin{equation}
\wt f_{ij} =  \delta_{ij}  f(\epsilon_i)
\end{equation}
This matrix is diagonal, however, its covariant derivative does not have to be diagonal (as we saw for the Hamiltonian in \eq{H-gender}.
Employing a Taylor expansion and a product rule we arrive at \eq{gender-chain-rule}, as shown in  
in Appendix \aref{chain-proof}.

%%% Local Variables:
%%% mode: latex
%%% TeX-master: "pap"
%%% End:

\section{Catalogue of geometric quantities}
\secl{geometric-quantity}

\tents{Before proceeding with derivation of explicit equations for Wannier interpolation of multiple geometrical quantities of interest,
we first define them in a gauge-covariant way and show that they all may be reduced to two objects $\Fcov^{ab}_{mn}$ and  $\Hcov^{ab}_{mn}$  defined below. }
%---------------------------------------------------
\subsection{Gauge-covariant matrices}
%---------------------------------------------------
\stm{the "covariant" definition here looks disconnected from previous 
section from first glance. Add some words to connect "P's" and "Q's" to
the definition of \eq{gender}}
Let $P=\sum_n^{\rm A}\,\ket{u_n}\bra{u_n}$ and $Q=1-P$.
% In the following, indices $m,n,n'$ run over the active space A.
The states $\ket{u_n}$ within the active space A are not assumed to be
energy eigenstates,\footnote{However, the A states $\{\ket{u_n}\}$ are
  assumed to be unitarily related to energy eigenstates.} so that in
general the Hamiltonian matrix
\beq
\Ham_{mn}\equiv\me{u_m}{\hat \Ham}{u_n}
\eeq
is not diagonal.
Following Eqs.~(6-8) of Ref.~\cite{lopez-prb12} We write the three gage-covariant quantities
\begin{align}
\wt F^{ab}_{mn}&\equiv\me{u_m}{(\partial_a P) Q (\partial_b P)}{u_n}=
\me{\partial_a u_m}{Q}{\partial_b u_n}\,,
\eql{Ftilde}\\
\wt G^{ab}_{mn}&\equiv
\frac{1}{2}\me{u_m}{\Ham (\partial_a P) Q (\partial_b P)+(\partial_a P) Q (\partial_b P)\Ham}{u_n}=\nonumber\\
&=\frac{1}{2}\left(\Ham_{mn'}\wt F^{ab}_{n'm}+
\wt F^{ab}_{mn'}\Ham_{n'n}\right)\,,
\eql{Gtilde}\\
\wt H^{ab}_{mn}&\equiv\me{u_m}{(\partial_a P)Q\Ham Q(\partial_b P)}{u_n}=
\me{\partial_a u_m}{Q\Ham Q}{\partial_b u_n}\,,
\eql{Htilde}
\end{align}
and also
\beq
\wt S^{ab}_{mn}\equiv\frac{\hbar^2}{m_e}\delta_{ab}\delta_{mn}+
2\left(\wt G^{ab}_{mn}-\wt H^{ab}_{mn}\right)\,.
\eql{Stilde}
\eeq
The four matrices $\wt O^{ab}_{mn}$ with $O=F,G,H,S$ are Hermitian
in the sence that $\wt O^{ab}_{mn} = \left(\wt O^{ba}_{nm}\right)^*$, 
and they (as well as $\Ham_{mn}$)
transform covariantly under unitary gauge transformations of the form
$\ket{u_n}\rightarrow\ket{u'_n}=\sum_m^{\rm A}\ket{u_m}U_{mn}$, that
is, $\wt O^{ab}\rightarrow U^\dagger \wt O^{ab} U$.

$\wt F^{ab}_{mn}$ is the metric-curvature tensor, from which the
covariant quantum metric and Berry curvature tensors
% $\wt{\cal F}^c_{mn}$ and
% $\wt f^{ab}_{mn}$
can be obtained as
\xlm{Change symbol of quantum matrix and inverse effective mass tensor from $f$ and $s$ to $\mathfrak F$ and $\mathfrak S$. $f$ is distribution function above.}
\beq
\wt{\mathfrak F}^{ab}_{mn}\equiv\frac{1}{2}\wt F^{ab}_{mn}+\frac{1}{2}\wt F^{ba}_{mn}
\eql{metric-cov}
\eeq
and
\beq
\wt{\Omega}^c_{mn}\equiv i\abc\wt F^{ab}_{mn}\,,
\eql{curv-cov}
\eeq
respectively. In the same way, we can extract from $\wt S^{ab}_{mn}$
two new covariant tensors
\beq
\wt {\mathfrak S}^{ab}_{mn}\equiv\frac{1}{2}\wt S^{ab}_{mn}+\frac{1}{2}\wt S^{ba}_{mn}
\eql{mass-cov}
\eeq
and
\beq
\wt{\cal S}^c_{mn}\equiv i\abc\wt S^{ab}_{mn}\,.
\eql{moment-cov}
\eeq
$\wt {\mathfrak S}^{ab}_{mn}$ is a generalized inverse effective mass tensor in a
sense to be clarified shortly, and $\wt{\cal S}^c_{mn}$ is a
generalized orbital magnetic moment tensor in the following sense: if
the A states are all degenerate with energy $\varepsilon_{\rm A}$, so
that $\Ham_{mn}=\varepsilon_{\rm A}\delta_{mn}$ in any gauge, the matrices
$\wt m^c_{mn}=(e/4\hbar)\wt{\cal S}^c_{mn}$ and
$\wt L^c_{mn}=-(m_e/2\hbar)\wt{\cal S}^c_{mn}$ reduce to the orbital
moment and orbital angular momentum matrices as defined in Eq.~(51) of
Ref.~\cite{chang-jpcm08},
\begin{multline}
\frac{2\hbar}{e}\wt m^c_{mn}=-\frac{\hbar}{m_e}\wt L^c_{mn}=
i\abc\left(\varepsilon_{\rm A}\wt F^{ab}_{mn}-\wt H^{ab}_{mn}\right)= \\=
i\abc\me{\partial_a u_m}{Q(\varepsilon_{\rm A}-\Ham)Q}{\partial_b u_n}\,.
\end{multline}

Finally, we also introduce a $\k$-resolved orbital magnetization
matrix
\begin{align}
\frac{2\hbar}{e}\wt M^c_{mn}\equiv -i\abc
\left(\wt H^{ab}_{mn}+\wt G^{ab}_{mn}-2\ef\wt F^{ab}_{mn}\right)=\\=
\frac{2\hbar}{e}\wt m^c_{mn}+
2i\abc\left(\ef\wt F^{ab}_{mn}-\wt G^{ab}_{mn}\right)\,.
\eql{Morb}
\end{align}

\subsection{Gauge-invariant traces}

% \subsection{General case}

Given a covariant matrix $\wt O_{mn}$ with $m,n\in{\rm A}$, we write
its (gauge-invariant) trace as
\beq
O_{\rm A}\equiv\Tr_{\rm A}\,\wt O=\sum_n^{\rm A}\,\wt O_{nn}\,.
\eql{trace-cov}
\eeq
Thus ${\mathfrak F}^{ab}_{\rm A}={\mathfrak F}^{ba}_{\rm A}$ and ${\Omega}^c_{\rm A}$ are the
net quantum metric and Berry curvature of the A states respectively,
$M^c_{\rm A}$ is their net $\k$-resolved orbital magnetization, and
${\mathfrak S}^{ab}_{\rm A}={\mathfrak S}^{ba}_{\rm A}$ is the sum of the inverse effective
masses (times $\hbar^2$) of the A band states,
\beq
{\mathfrak S}^{ab}_{\rm A}:=\sum_n^{\rm A}\,\partial^2_{ab}\en\,.
\eeq
Here the symbol $:=$ denotes an equality whose right-hand-side
only holds in a gauge where the Hamiltonian matrix is diagonal:
$\Ham_{mn}=\varepsilon_m\delta_{mn}$.\footnote{Note that if $X:=Y$ and
  $Z:=Y$, then $X=Z$.}

% \subsection{Single-band limit and band additivity}

If space A contains a single band $n$, we simplify the notation as
$\wt O^{ab}_{mn}\rightarrow O^{ab}_n$. In that limit
\eqr{Ftilde}{Htilde} reduce to
\begin{align}
F^{ab}_n&=\ip{\partial_a u_n}{\partial_b u _n}-\ip{\partial_a u _n}{u_n}\ip{u_n}{\partial_b u _n}\,,\\
G^{ab}_n&=\en F^{ab}_n\,,\\
H^{ab}_n&=\me{\partial_a u _n}{\Ham}{\partial_b u _n}-
\en\ip{\partial_a u _n}{u_n}\ip{u_n}{\partial_b u _n}\,,
\end{align}
and \eq{Stilde} becomes
\beq
S^{ab}_n=\frac{\hbar^2}{m_e}\delta_{ab}+
2\me{\partial_a u _n}{(\en-\Ham)}{\partial_b u _n}\,,
\eeq
which agrees with the expression given in Ref.~\cite{gao-prb15} (see
the 2nd column of p.~3 therein).

In the same limit \eqs{metric-cov}{curv-cov} reduce to the single-band
quantum metric and Berry curvature, respectively,
\begin{align}
{\mathfrak F}^{ab}_n&=\Re\,F^{ab}_n=
\Re\ip{\partial_a u _n}{\partial_b u _n}-\ip{\partial_a u _n}{u_n}\ip{\partial_b u _n}{u_n}
\,,\\
{\Omega}^c_n&=-\abc\Im\,F^{ab}_n=
% -\abc\Im\,\ip{\partial_a u _n}{\partial_b u _n}\,,
-\Im\,\bra{\nabla_\k u_n}\times\ket{\nabla_\k u_n}_c\,,
\end{align}
\eqs{mass-cov}{moment-cov} become proportional to the single-band
inverse effective mass and orbital moment, respectively,
\begin{multline}
{\mathfrak S}^{ab}_n=\Re\,S^{ab}_n=\\=
\frac{\hbar^2}{m_e}\delta_{ab}+2\Re\me{\partial_a u _n}{(\en-\Ham)}{\partial_b u _n}
=\partial^2_{ab}\en\,,
\end{multline}
\begin{multline}
{\cal S}^c_n =-\abc\Im\,S^{ab}_n=\\=
%-2\abc\Im\me{\partial_a u _n}{(\epsilon_n-\Ham)}{\partial_b u _n}=
-2\Im\bra{\nabla_\k u_n}\times (\en-\Ham)\ket{\nabla_\k u_n}_c=
\frac{4\hbar}{e}m^c_n\,,
\end{multline}
and \eq{Morb} becomes the single-band $\k$-resolved orbital
magnetization,
\beq
\frac{2\hbar}{e}M^c_n=\Im\bra{\nabla_\k}\times
\left(\Ham+\en-2\varepsilon_{\rm F}\right)\ket{\nabla_\k u_n}_c\,,
\eeq
or equivalently,
\beq
M^c_n=m^c_n+\frac{e}{\hbar}(\varepsilon_{\rm F}-\en){\Omega}^c_n\,.
\eeq

Suppose that space A is entirely made up of bands that never touch one
another. Then its net Berry curvature, inverse effective mass, and
orbital magnetization are  equal to the sums over bands of the
corresponding single-band quantities,
%
%\begin{align}
%{\Omega}^c_{\rm A}&=\sum_n^{\rm A}\,\Omega^c_n\,,\\
%{\mathfrak S}^{ab}_{\rm A}&=\sum_n^{\rm A}\,{\mathfrak S}^{ab}_n\,,\\
%M^c_{\rm A}&=\sum_n^{\rm A}\,M^c_n\,,
%\end{align}
\beq
{\Omega}^c_{\rm A}=\sum_n^{\rm A}\,\Omega^c_n\,,\quad
{\mathfrak S}^{ab}_{\rm A}=\sum_n^{\rm A}\,{\mathfrak S}^{ab}_n\,,\quad
M^c_{\rm A} =\sum_n^{\rm A}\,M^c_n\,,
\eeq
and such quantities are said to be ``band additive.'' Note that the
quantum metric and the orbital moment are {\it not} band additive,
since in general
\beq
{\mathfrak F}^{ab}_{\rm A}\not=\sum_n^{\rm A}\,{\mathfrak F}^{ab}_n\,,\quad
{\cal S}^c_{\rm A}\not=\sum_n^{\rm A}\,{\cal S}^c_n\,.
\eeq

%%% Local Variables:
%%% mode: latex
%%% TeX-master: "pap"
%%% End:

\section{Wannier interpolation}

\secl{wannier-interpolation}
%\section{Berry curvature and its gradient in the minimal tight-binding formalism} 
\subsection{Wannier functions and effective models}
\secl{Wannier}
In this section we introduce the necessary notation, and 
briefly recall the spirint of Wannier interpolation of Berry curvature, 
closely following Ref.~\cite{wang-prb06}.

Wannier functions $\ket{\RR j}$ form a localized orthonormal basis for the description of electron bandstructure.\cite{Marzari-MLWF-rmp}
The eigenvalues of the Wannier Hamiltonian 
\beq
\Ham\w_{ij}(\kk)=\sum_\R\,e^{i\k\cdot(\R+\bt_j-\bt_i)}\Ham\w_{ij}(\R)\,.
\eql{H-k}
\eeq
accurately reproduce the eigenvalues of the Bloch bands computed from first principles. 
Here $\RR$ are lattice vectors, and the matrix elements $\Ham\w_{ij}(\R)$ are computed as 
\beq
\Ham\w_{ij}(\RR)=\me{\R' i}{\hat{\Ham}}{\R'+\R,j}=
\me{{\bf 0}i}{\hat \Ham}{\R j}\,,
\eql{H-R}
\eeq
and the orthonormality condition reads
\beq
\ip{\RR'i}{\RR j} = \delta_{\RR\RR'} \delta_{ij} \eql{orthonormal}
\eeq
The electron energies and wavefunctions
at any arbitrary wavevector $\kk$ are obtained from the secular equation
\beq
\Ham\w(\k) U\bnk=\enk U\bnk\,,
\eql{eig-wann}
\eeq
to find the eigenenergies $\enk$ and  column vectors $\ket{ n}$
of coefficients $U_{jn}(\kk)$  of the expansion of the wavefunctions
\beq
\ket{u_{j\kk}}=\ket{u\ww_{j'\kk}} U_{j'j}(\kk)
\eeq
in the Bloch basis $\ket{u\ww_{j\kk}}$  constructed from Wannier functions as
\beq
\ket{u\ww_{j\kk}} = \sum_\RR e^{i\kk\cdot(\RR+\bt_j-\rr)}\ket {\RR j}\,.
\eeq
We have included Wannier centers $\bt_i$ in the phase factors
the phase factor, as this is the most convenient convention for
handling Berry-phase quantities~\cite{vanderbilt-book18}.(See \aref{wannier-centers} for details )

The derivative of states $\ket{n}$ of the Wannier Hamiltonian is given by \eq{dn-1} 
the and taking inner product with $\bra{ l}$ yields the anti-Hermitian matrix
\beq
D^a_{ln}=\ip{l}{\partial_a n}=
\begin{cases}
\dis
\frac{\lme{l}{\partial_a \Ham\w}{n}}{\en-\el},& \text{if $l\not= n$}\\
-i\alpha_n,& \text{if $l=n$}
\end{cases}\,.
\eql{D-a}
\eeq
which is convenient for the evaluation of the derivative of Bloch states as 
\beq
\ket{\partial_b u_n} =  \ket{\partial_b u_{j'}\ww}U_{j'n} +  \ket{u_{j'}}D_{{j'}n}^b
\eql{der-u}
\eeq
Inserting \eq{der-u} into \eq{curv-n-def} and summing over states in set A one gets the net Berry curvature
\beas{berry-curv-wannier-int}
\Omega^{ab}_\mathrm{A} &=& \Omega^{ab}_\mathrm{A,int}+ \Omega^{ab}_\mathrm{A,ext}~;\\
\Omega^{ab}_\mathrm{A,int}&=& 2\Im\sum^A_n\sum^B_l D^a_{nl} D^b_{ln}~;
\eql{Omega-DD}\\
\Omega^{ab}_\mathrm{A,ext} &=& 
 -2\Im \sum^A_n \overline{F}^{ab}_{nn}
-4 \Re \sum^A_n\sum^B_l  D^a_{nl} \overline{A}^b_{ln}~,
\eql{Omega-DD-AD}
\eeas
which we have separated into ``internal'' and ``external'' terms for a reason that will be explained shortly.
Hereinafter, with an overline we denote a transform of any matrix object in the Wannier gauge ${\cal O}\w_\k$ 
into to the Hamiltonian gauge: 
\beq
\overline{\cal O}_{mn} = \me{m}{{\cal O}\w_\k}{n}= \left( U^\dagger \cdot {\cal O} \cdot U\right)_{mn}
\eql{O-mn}
\eeq
and the corresponding Wannier gauge matrices $A\w$ and $F\w$ are defined in \eqs{Aww-2}{Fww-2} based on the real-space matrix elements \eqref{eq:O-R}.

In the derivations above, one could replace the Wannier functions to any other set  of orthonormal
\footnote{Generalization to non-orthogonal localized basis was done in \citep{Wang_2019-nonorthogonal,Jin_2021-nonorthogonal}.
Our formalism of covariant derivatives can also be generalized to that case, but we leave it out of the scope of the present article.
}
 localized basis states $\ket{\RR j}$. In fact, the derivation would be the same for an empirical tight-binding model, with the only difference 
 that the matrix elements $\Ham_{ij}(\RR)$, $\mathbb{A}^a_{ij}(\RR)$ and $\mathbb{F}^a_{ij}(\RR)$ would not be computed via \eqs{H-R}{O-R}, but rather fitted to bandstructure or chosen empirically. It is a common practice in tight-binding models to neglect matrix elements \eqref{eq:O-R}, and work only with the ``hoppings'' $\Ham(\RR)$.  Also, when working with an effective $\kk\cdot\mathbf{p}$ model, there are no localized functions, 
 but instead the Hamiltonian matrix is assumed to be written in a basis that does not  depend on $\kk$, and therefore the first term in \eq{der-u} vanishes. In these cases the Berry curvature is given only by $\Omega^{ab}_\mathrm{A,int}$ which is identical to \eq{Omega-A-1}. 
The terms that survive in the case of an effective model are called \textit{internal} because they are the property of a Hamiltonian. In turn, the terms that depend on additional matrix elements are called \textit{external}, and they are important for an accurate \textit{ab initio} description of
electronic properties.  In the rest of the present article we will keep this separation.

%%% Local Variables:
%%% mode: latex
%%% TeX-master: "pap"
%%% End:

\subsection{Wannier interpolation of 
$\Fcov^{ab}_{mn}$ and  $\Hcov^{ab}_{mn}$ }
\secl{wannier-interpolation-FH}

\ism{In case we discuss internal vs external terms of the orbital
  moment matrix, we should try to make contact with the $g$-factors
  literature, where a similar distinction is made: see Eq.~(19) in
  Phys. Rev. Research {\bf 2}, 033256 (2020). \tenxl{ Do you mean Eq.~18 ?}}

In this section we derive the Wannier interpolation of quantities given by \eqs{Ftilde}{Htilde}.
% the follwoing covariant quantities:
%which we rewrite as 
%%%
%\bea
%\Fcov^{ab}_{mn}&=\ip{\wt\partial_a u_m}{\wt\partial_b u_n} &=\left(\Fcov^{ba}_{nm}\right)^*\,, \eql{Fab}\\
%\Hcov^{ab}_{mn}&=\me{\wt\partial_a u_m}{\Ham}{\wt\partial_b u_n}&=\left(\Hcov^{ba}_{nm}\right)^*\,, \eql{Hab}
%\eea
%where $\wt\partial_a \equiv Q\partial_a$, and recall that 
%%
%The approach closely follows Ref. \cite{wang-prb06}. 
%And let us consider Bloch states in another gauge
%\beq
%\ket{u_{j\kk}}=\ket{u\ww_{n\kk}} U_{j'j}(\kk)
%\eeq
%These may be the eigenstate of the Hamiltonian (Hamiltonian gauge) or any other 
%gauge, that is related to the hamiltonian gauge by a transformation not mixing 
%the subspaces $A$ and $B$. 
%According to eq. 26 of  \cite{wang-prb06} the derivative with respect to $\kk$ vector may be taken as
%%
%\beq
%\ket{\partial_b u_n} =  \ket{\partial_b u_{j'}\ww}U_{j'n} +  \ket{u_{j'}}D_{{j'}n}^b
%\eeq
Recalling that  $Q=1-\ket{u_{n'}}\bra{u_{n'}}$  from \eq{der-u} we obtain
\begin{multline}
Q\ket{ \partial_b u_n} = Q \ket{\partial_b u_{j'}\ww}U_{{j'}n} +  \ket{u_{l'}}D_{l'n}^b =  \\
\ket{\partial_b u_{j'}\ww} U_{j'n}
- \ket{u_{n'}} \ip{u_{n'}}{\partial_b u_{j'}\ww}U_{j'n}
+   \ket{u_{l'}}D_{l'n}^b=
\\
\ket{\partial_b u_{j'}\ww}U_{j'n} + i\sum_{n'}\inn \ket{u_{n'}} \overline{A}^b_{n'n}+ \ket{u_{l'}}D_{l'n}^b
\end{multline}
and using and $Q^2=Q$ we obtain the following expressions:
\beas{Fab-wanint-all}
\Fcov^{ab}_{mn}              &=& \Fcov^{ab}_{mn,\textrm{int}} + \Fcov^{ab}_{mn,\textrm{ext}} \eql{Fab-wanint}\\
\Fcov^{ab}_{mn,\textrm{int}} &=& - D_{nl'}^a D_{l'm}^b \eql{Fab-wanint-int}\\
\Fcov^{ab}_{mn,\textrm{ext}} &=&  
\left[\left(iD^a_{nl'}\Abar^b_{l'm}\right)+\conjabmn   \right]
\nonumber \\ &&
+\Fbar^{ab}_{mn}-  \Abar^a_{mn'} \Abar^b_{n'n} 
\eql{Fab-wanint-ext}
\eeas
and
\beas{Hab-wanint-all}
\Hcov^{ab}_{mn}&=& \Hcov^{ab}_{mn,\textrm{int}} + \Hcov^{ab}_{mn,\textrm{ext}} \eql{Hab-wanint}\\
\Hcov^{ab}_{mn,\textrm{int}}&=&- D_{nl'}^a \Ham_{l'l''} D_{l''m}^b  \eql{Hab-wanint-int}\\
\Hcov^{ab}_{mn,\textrm{ext}}&=& 
 \left[\left(i D^a_{nl'}\overline{B}^b_{l'm}\right)+\conjabmn \right] +
\nonumber\\&&
+\Hbar^{ab}_{mn} -  \Abar^a_{mm'} \Ham_{m'n'} \Abar^b_{n'n} 
\eql{Cab-wanint}
\eeas
Similar to \eq{berry-curv-wannier-int} we have separated the result into internal and external terms. 
Here the bar above a matrix follows \eq{O-mn}.
%\beq
%\overline{\cal O}_{ij}=\left[ U^\dagger{\cal O}\ww U\right]_{ij}  \,,
%\eql{O-bar}
%\eeq
To describe the external terms, And following \cite{wang-prb06} and \cite{lopez-prb12} 
we have defined a series of quantities defined in the Wannier gauge:\footnote{Note that in \cite{lopez-prb12} the quantity of our \eq{Hww} was denoted as $C\ww$}
\beas{Oww-all}
\left(A\ww\right)^{a}_{ij}&\equiv&  i\ip{ u_i\ww}{\partial_a u_j\ww} \eql{Aww} \\% = \nonumber\\ 
%   &=& \sum_\mathbf{R} e^{i\kk(\R+\bt_j-\bt_i)}\memnR{\hat{r}_a-\RR-\bt_j } =
%      e^{i\kk(\bt_j-\bt_i)}\sum_\mathbf{R} e^{i\kk\R}  \mathbb{A}^a_{ij}(\RR)  \\
\left(B\ww\right)^{a}_{ij}&\equiv&  i\me{ u_i\ww}{\Ham}{\partial_a u_j\ww} \eql{Bww}\\
%  =  \sum_\mathbf{R} e^{i\kk\mathbf{R}}\mathbb{B}^a_{ij}(\RR)   \\
\Hwan^{ab}_{ij}&\equiv&  \me{\partial_a u_i\ww}{\Ham}{\partial_b u_j\ww} \eql{Hww} \\
% = \sum_\mathbf{R} e^{i\kk\mathbf{R}}\mathbb{C}^{ab}_{ij}(\RR) \\
\Fwan^{ab}_{ij}&\equiv&  \ip{\partial_a u_i\ww}{\partial_b u_j\ww}  \eql{Fww}
 % = \sum_\mathbf{R} e^{i\kk\mathbf{R}} \mathbb{F}^{ab}_{ij}(\RR)\\
\eeas
Evaluation of these quantities is given in \aref{wannier-gauge}

Inserting \Eqs{Fab-wanint-all}{Hab-wanint-all} into corresponding equations from \sref{geometric-quantity} one can obtain Wannier interpolation of the needed geometrical quantity. For instance, combining \eqref{eq:Fab-wanint-all}, \eqref{eq:curv-cov} and \eqref{eq:trace-cov} one obtains the known result for Berry curvature \eqref{eq:berry-curv-wannier-int}.

\subsection{Covariant  derivatives of %  
$\Fcov^{ab}_{mn}$ and $\Hcov^{ab}_{mn}$}
\secl{der-F-H}

Now we are ready to take the covariant derivatives of \eqs{Fab-wanint}{Cab-wanint}, using the product rule defined in \eq{gender-prod-rule}.

\beas{Fab-wanint-der}
\Fcov^{ab:d}_{mn}&=& \Fcov^{ab:d}_{mn,\textrm{int}} + \Fcov^{ab:d}_{mn,\textrm{ext}} \eql{Fab-wanint-der-all}\\
\Fcov^{ab:d}_{mn,\textrm{int}}&=& D_{ml'}^{a} D_{l'n}^{b:d} +\conjabmn \eql{Fab-wanint-der-int}\\
\Fcov^{ab:d}_{mn,\textrm{ext}}&=& - \Bigl[\Bigl( i\Abar^a_{ml'}D^{b:d}_{l'n} + iD^a_{ml'}\Abar^{b:d}_{l'n}\nonumber \\
 & &  +\Abar_{mn'}^a\Abar_{n'n}^{b:d} \Bigr) +\conjabmn \Bigr]+ \Fbar^{ab:d}_{mn}   \eql{Fab-wanint-der-ext}
\eeas
\beas{Hab-wanint-der}
\Hcov^{ab:d}_{mn}&=& \Hcov^{ab:d}_{mn,\textrm{int}} + \Hcov^{ab:d}_{mn,\textrm{ext}} \eql{Hab-wanint-der-all}\\
\Hcov^{ab:d}_{mn,\textrm{int}}&=&
			        - \Bigl[ D_{ml'}^{a} {\Ham_{l'l''}} D_{l''n}^{b:d}  +\conjabmn \Bigr] 
			\nonumber\\ &&
			-  D_{ml'}^{a} \Ham_{l'l''}^{,d} D_{l''n}^b  			        
			        \eql{Hab-wanint-der-int}\\
\Hcov^{ab:d}_{mn,\textrm{ext}}&=& \Bigl[
                 i(\overline{B}^\dagger)^a_{ml'}D^{b:d}_{l'n} + iD^a_{ml'}\overline{B}^{b:d}_{l'n}  \nonumber\\&&
        -  \Abar_{mm'}^a \Ham_{m'n'} \Abar_{n'n}^{b:d}  
         +\conjabmn  \Bigr]  -  
                      \nonumber\\ &&
                 -\Abar_{mm'}^a \Ham_{m'n'}^{,d} \Abar_{n'n}^{b}  + \Hbar^{ab:d}_{mn} 
\eql{Hab-wanint-der-ext}
\eeas
Now we need to find the generalized derivatives of the ingredients of this equation. 

The derivatives $\Abar^{b:d}_{ln}$, $\overline{B}^{b:d}_{ln}$, 
$\overline{C}^{b:d}_{ln}$ and $\overline{F}^{b:d}_{ln}$ are evaluated in a general way.  
Consider a quantity of the form of \eq{O-mn}, \eq{gender-Xnn} for
instance reads
\beq
\overline{\cal O}^{:d}_{nn'}=\overline{\cal O}^{,d}_{nn'}
-\sum_l\out D^d_{nl}{\cal O}_{ln'}+\sum_l\out {\cal O}_{nl}D^d_{ln'}\eql{gender-Onn}
\eeq
where in accordance with \eq{operator-coma-derivative}  we defined
\beq
\overline{\cal O}^{,d} \equiv\overline{\partial_d {\cal O}}=
U^\dagger\left(\partial_d {\cal O}\ww\right) U\,. \eql{commader}
\eeq
Some technical notes on the evaluation on the evaluation of $\overline{B}^{b:d}_{ln}$ are left for \aref{B-frozen}.
In order to derive $D_{ln}^{b:d}$  let's rewrite \eq{D-a} for $D^b_{ln}$ as 
\beq
  D_{ln'}^b \Ham_{n'n} -  \Ham_{ll'}D_{l'n}^b = \Ham_{ln}^{,b}~.
 \label{eq:DHmHD}
\eeq
Now we take the covariant derivative of
both sides of this equation and employing the product rule
\eq{gender-prod-rule} we get
\begin{multline}
  D_{ln'}^b \Ham_{n'n}^{:d}  + D_{ln'}^{b:d}  \Ham_{n'n} 
- \Ham_{ll'}D_{l'n}^{b:d}  -  \Ham_{ll'}^{:d}D_{l'n}^b =\\
 \Ham_{ln}^{,bd} - D_{ln'}^{d} \Ham_{n'n}^{,b} +  \Ham_{ll'}^{,b} D_{l'n}^d 
\end{multline}
where we also used \eq{gender-Xln} for $\Ham_{ln}^{,b:d}$ and
\eq{H-gender}

Now, consider this equation in the Hamiltonian gauge,
collect all terms except $D_{nl'}^{b:d}$ in the RHS and
divide by $(\en-\el)\equiv \varepsilon_{nl}$ to get the following  expression 
\beq
D^{b:d}_{ln}=\frac{1}{\varepsilon_{nl}}
\left[
\Ham^{,bd}_{ln}+
\left(\Ham^{,b}_{ll'} D^d_{l'n} - D^d_{ln'}\Ham^{,b}_{n'n}  + (b \leftrightarrow d)\right)
\right]
\eql{gender-sumrule}
\eeq

It can be seen that the derived equations do not contain the $D$ matrix states A, as well as between states B.  As far as spaces A and B are 
By construction, this behaviour will be preserved in evaluation of higher derivatives (see \aref{second-derivative}).

Inserting \Eqs{Fab-wanint-der-all}{Hab-wanint-der-all} into corresponding equations from \sref{geometric-quantity} one can obtain Wannier interpolation of thederivatives of the needed geometrical quantities. For instance, combining \eqref{eq:Fab-wanint-der-int}, \eqref{eq:der-trace}, \eqref{eq:curv-cov} and \eqref{eq:gender-sumrule} one obtains the known result for derivative of the Berry curvature of occupied states \eqref{eq:Omega-A-grad-1}, that we gave in \sref{preliminary} without proof.

%%% Local Variables:
%%% mode: latex
%%% TeX-master: "pap"
%%% End:

\section{First principles results: trigonal Tellurium}
\secl{result-Te}

%%% Local Variables:
%%% mode: latex
%%% TeX-master: "pap"
%%% End:TR
Trigonal Tellurium has a crystal structure composed of homotropic 
triple helix Te-chains with space group $P3_121$ (right handed) or 
$P3_221$ (left handed), the right handed structure of Te is shown in \fref{Te-Cell}. 
The three Te atoms in each unit cell are evenly distributed along the helix. 
The screw structure breaks inversion symmetry and creates a Berry curvature dipole.
\begin{figure}[t]
\begin{center}
\includegraphics[width=0.9\columnwidth]{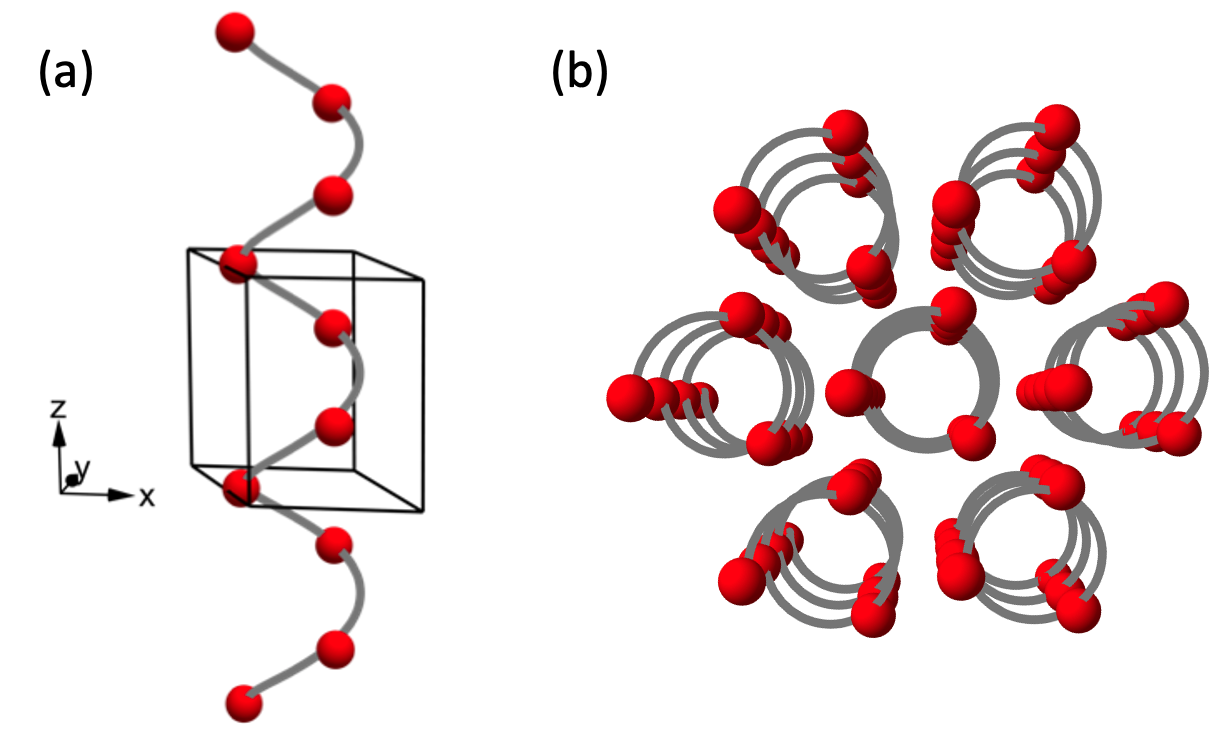}
    \caption{Crystal structure of tellurium with space group $P3_121$.(a) The tellurium chains spiral along z-axes of the crystal lattice. (b) Top view of Te spiral chains.}
    \figl{Te-Cell}
\end{center}
\end{figure}

The electronic structure is calculated by employing the HSE06 hybrid functional \cite{paier2006screened} implemented in the VASP code \cite{Kresse:1996,kresse1996efficiency,Kresse:1999} .
The maximaly localized Wannier functions are generated using Wannier90 \cite{Pizzi:2020}, and disentanglement is performed using a frozen window below $ \varepsilon_{\rm F} + 3{\rm eV} $ with $s, p$ orbitals of Te as projections. 
Tellurium has a band gap of about 0.3eV with conduction band (CB) minimum and valence band (VB) maximum around the H and H$^\prime$ points of the brillouin zone (BZ).
The material is usually p-doped, and the VB presents more interest. However, for our demonstration the CB is more interesting because there is a Weyl point (WP) at the H point, which presents a computational challenge for the evaluation of 
Berry dipole via \eq{dp-surf}. Moreover, this WP is predicted to give a significant contribution to switch the sign of circular photogalvanic effect at high temperatures and low doping \cite{tsirkin-gyrotropic}.
The methodology derived in this manuscript has been implemented within the open-source code WanierBerri\cite{wannierberri}. 
% There are Weyl points located at the $H$ or $H'$ points with energy 0.312eV above the Fermi level. 
%All Fermi surfaces at the energy of the Weyl point are contributed by Rashba-splitting-type band structures around the Weyl points, as shown in \fref{single-band-compare-Te}(c). 
\begin{figure}
\begin{center}
\includegraphics[width=1.0\columnwidth]{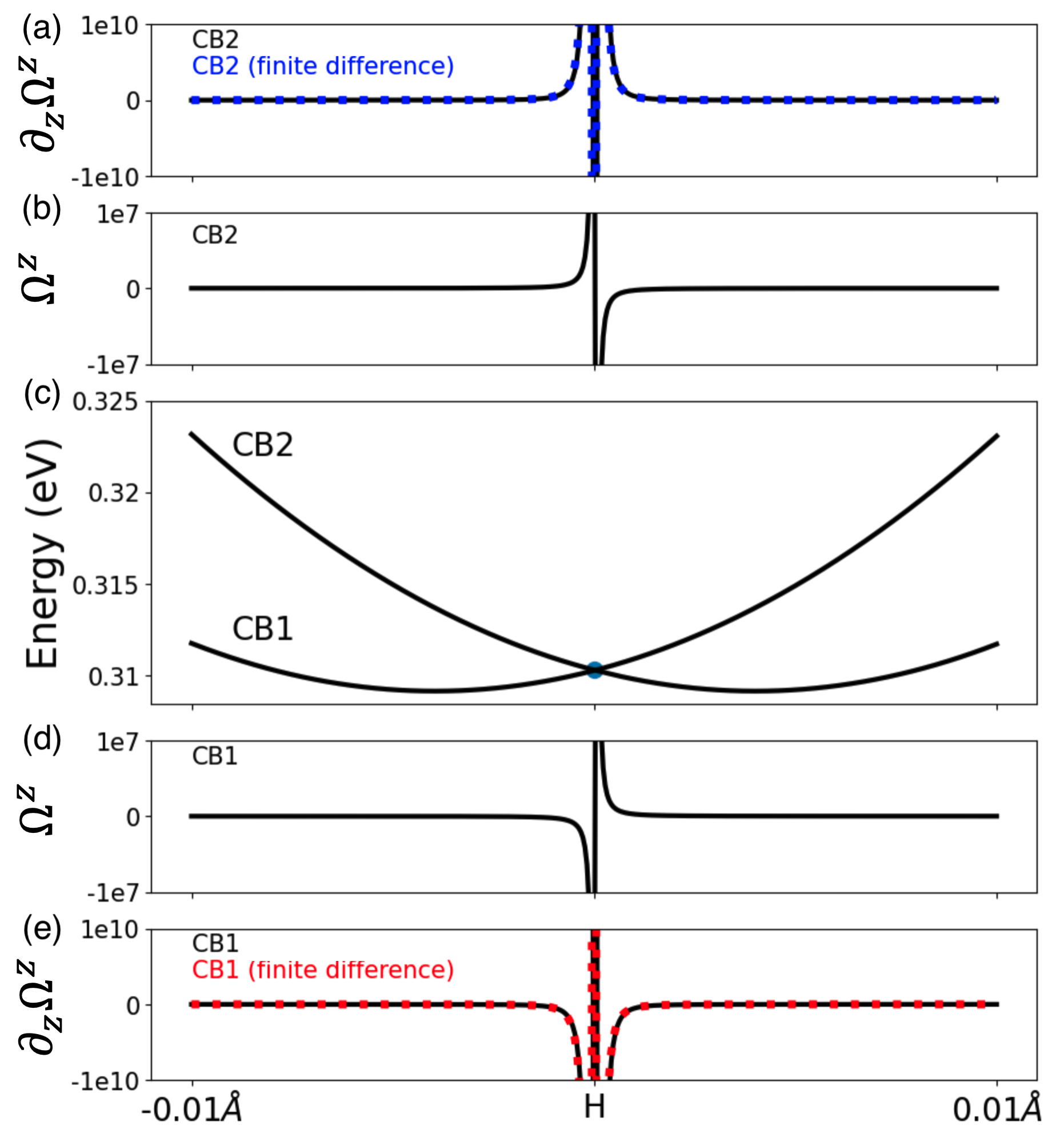}
\caption{(c)  Energy band of Tellurium, the k-path of 
    which is a small range around H point of K-H-K line. 
    There is a Weyl point (WP) at the H point located at 0.312eV 
    above Fermi energy. (b)(d) Berry curvature z-component 
    of the conduction band (CB) 1 and 2. (a)(e) Derivative 
    of Berry curvature z-component of CB1 and CB2. Solid black 
    lines are calculated with \eqs{Fab-wanint-der-int}{Fab-wanint-der-ext}. 
    And the dashed color lines are calculated by the finite difference of Berry curvature.}
\figl{single-band-compare-Te}
\end{center}
\end{figure}
By using Wannier interpolation, 
the interpolated Berry curvature $\wt \Omega$ of conduction bands CB1 and CB2  can be calculated with
\beq
\wt \Omega^c_n = -\epsilon_{abc} {\rm Im} \wt F^{ab}_{nn}~,
\eeq
where $\wt F^{ab}_{nn}$ is introduced in \eq{Fab-wanint}.
According to \eq{Omega-kn}, the Berry curvature of a single band follows an inverse-square law with respect to the energy difference with other bands,
and therefore it changes rapidly in the vicinity of the WP, as shown in \fref{single-band-compare-Te}(b,d). 
Following \eq{Fab-wanint-der-all}, the interpolated derivatives of Berry curvature are calculated with
\beq
\wt \Omega^{c:d}_n = -\epsilon_{abc} {\rm Im} \wt F^{ab:d}_{nn},
\eeq
which are shown in \fref{single-band-compare-Te}(a,e) as solid lines. 
The dashed colored lines are plotted using finite differences based on the interpolated Berry curvature data in \fref{single-band-compare-Te}(b,d), 
using denser k sampling. The solid curves in \fref{single-band-compare-Te}(b,d) show good agreement with dashed colored lines, 
indicating that our Wannier interpolation method with covariant gradient works well around WPs in real materials.
\begin{figure}[t]
\begin{center}
\includegraphics[width=0.9\columnwidth]{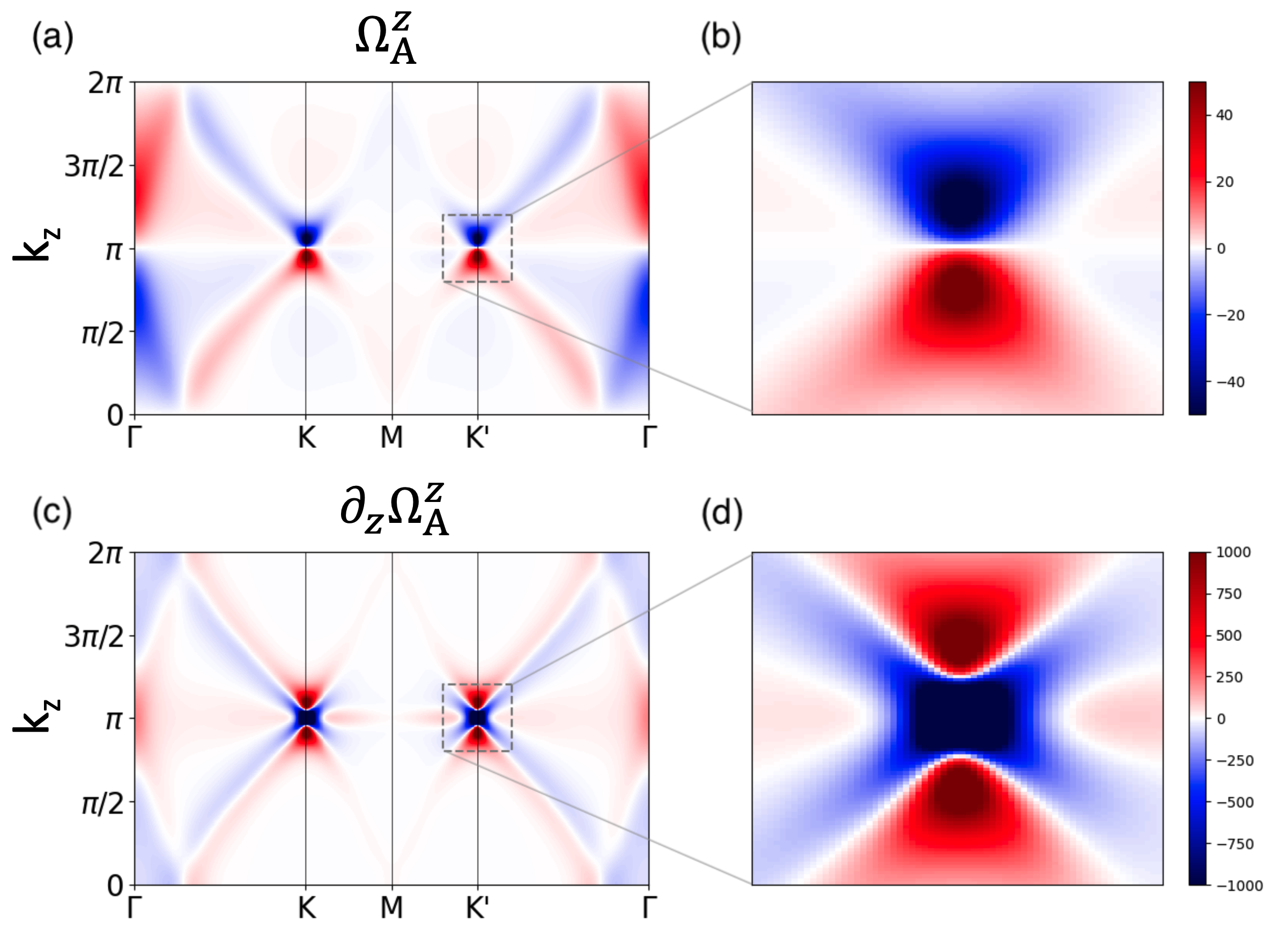}
    \caption{(a)(c) Berry curvature z-component $\Omega^z_{\rm A}$ and 
    derivative of Berry curvature zz-component  $\partial_z\Omega^z_{\rm A}$ 
    of all occupied bands below Weyl point (WP) energy on a plane submanifold 
    of Brillouin zone. (b)(d) The zoom-in of the dashed square range in (a)(c). }
\figl{plane-Te-dp}
\end{center}
\end{figure}

%Due to time-reversal symmetry, the berry curvature and its gradient obey the following relation
%\beas{Omega-k}
%\wt \Omega^z_{\k,{\rm n}} &= -\wt \Omega^z_{-\k,{\rm n}} \eql{Omega+-k}\\
%\partial_z\wt\Omega^z_{\k,{\rm n}} &= \partial_z\wt\Omega^z_{-\k,{\rm n}} \eql{dOmega+-k}
%\eeas
%shown in \fref{plane-Te-dp}(a). 
In the previous research, 
evaluating  $ \Ham_{\k n}^{:a} \wt \Omega_{\k n}^b$ on the Fermi surface is 
a widely used method to calculate dimensionless Berry curvature 
dipole tensor via \eq{dp-surf}. 
Typically, one evaluates it as a direct summation 
\beq
\mathcal{D}_{ab}^{\rm surf} = \frac{1}{N_\kk V_{\rm u.c.}}\sum_{\kk n}  \partial_a \enk \Omega_{\k n}^b \left(- \frac{\partial f_0}{\partial \varepsilon}\right)_{\varepsilon = \enk}~.
\eql{dp-surf}
\eeq
However, at low temperature, the derivative of the distribution function with energy $f_0'$ 
is a narrow peak function. This means only the k points that lie closely on the Fermi surface contribute to the integral.
However, a big number of points needs to be evaluated before the narrow peaks from individual $\kk$-points merge into a smooth curve. 
In \cite{Singh-MoTe}  $\mathcal{D}_{ab}$ was by first computing the Fermi surface by employing the
tetrahedron method at a given k grid, and then the Berry curvature was sampled only at the reduced grid points near the Fermi
surface. However, they also noted a slow convergence of the integral. 

Instead, in \eqref{eq:dp-sea} all k-points (which have bands below Fermi level) contribute, and therefore we are integrating a smoother function. 
%By using integration by parts, the derivative can be moved from the distribution function $f_0$ to Berry curvature $\Omega_{\k n}$ \eq{dp-sea}
%\beq
%\mathcal{D}_{ab}^{\rm sea} = \int \dk \sum_n \wt \Omega_{\k n}^{b:a} f_0(\enk)~,
%\eql{dp-sea}
%\eeq
%And the evaluation method involves a Fermi sea integral, which requires summing up the
%\tenxl{$\Omega^{b:a}_{\k n}$} values of all occupied bands over the entire first Brillouin zone. 
%All points on the k-grid are used in the calculation of these quantities. 
 %, which relies on only a few k-points.
%By using the Wannier interpolation method with covariant gradient, we can interpolate the matrix $\wt \Omega_{\k n}^{b:a}$ 
%on the k-grid to produce smooth results, in contrast to the jagged data obtained from finite differences.
%And this interpolation method aids in achieving convergence.

Another reason for the slow convergence of \eq{dp-surf}
is the sharp peak of the Berry curvature near a WP.  Close to theWP the divergent part of the BC is equal in magnitude and opposite in sign for the two subbands. Therefore, if the Fermi level is above (below) both subbands, in the Fermi-sea integral the divergent parts cancel out (do not appear). In turn, in the Fermi-surface integral both subbands contribute with their divergent Berry curvature multiplied by different velocities and distribution functions, and therefore no cancellation occurs.

\begin{figure}[t]
\begin{center}
\includegraphics[width=0.95\columnwidth]{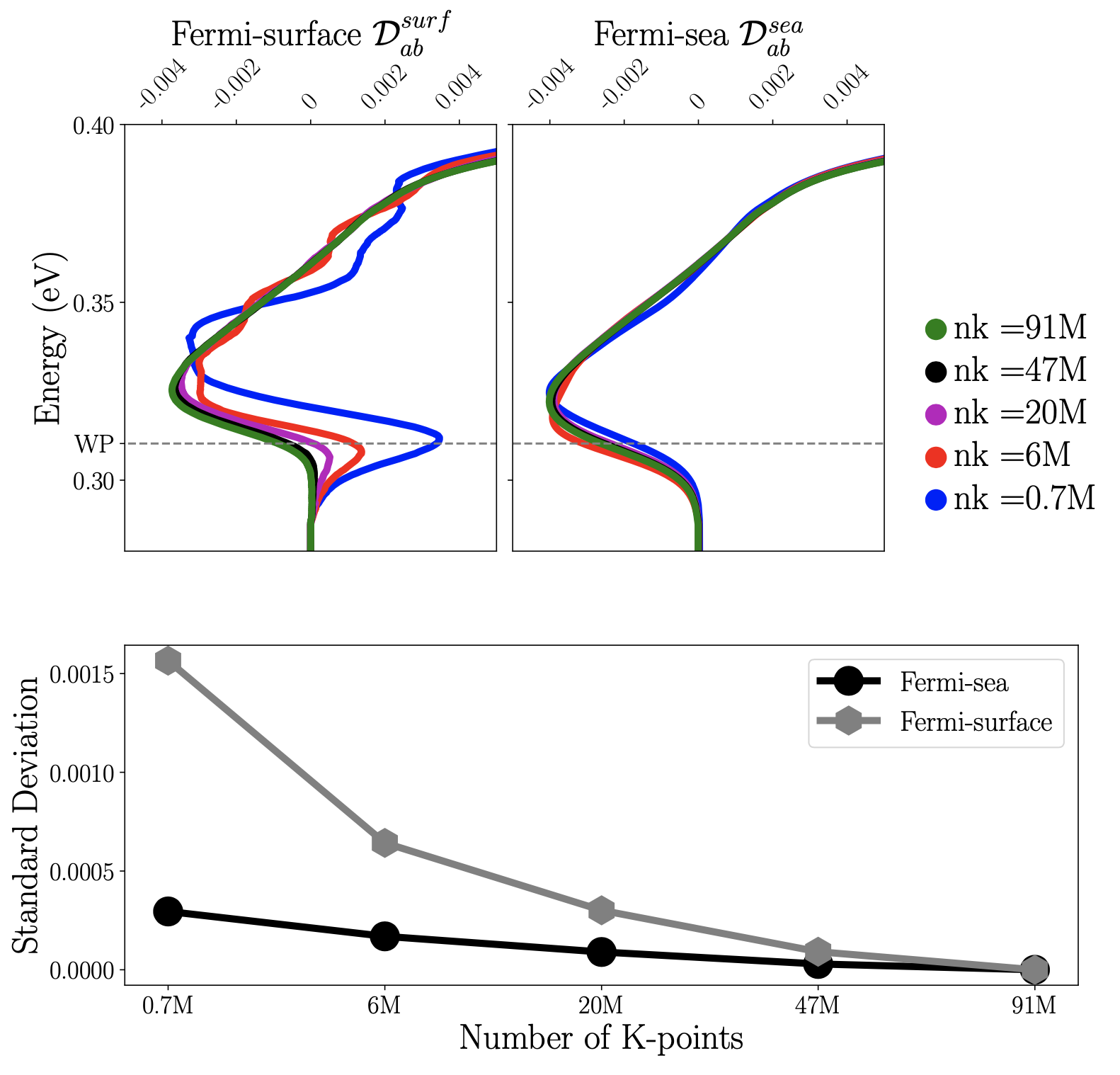}
\caption{(a) Integral of Berry curvature dipole using Fermi surface 
    \eq{dp-surf} and Fermi sea \eq{dp-sea} integral function. 
    The colors show different k-points sampling numbers (unit M is million) 
    in all Brillouin zone. WP is the energy where the Weyl point 
    is located. (b) The standard deviation compared with the converged result. }
\figl{fsur-sea-dipole}
\end{center}
\end{figure}

Thus, the Fermi-sea integral should have a better convergence with respect to the density of the $\kk$-grid.
To demonstrate this, we calculated the Berry dipole using both \eq{dp-surf} and \eq{dp-sea} at temperature 50K, as shown in \fref{fsur-sea-dipole}(a),
with varying numbers of k-points.
When using \eq{dp-surf} and a smaller number of k-points, 
there is a clear divergence in the Fermi-surface curve at the Weyl point energy.
And the curve does not converge until 47 million k-points are used.
In contrast, the Fermi-sea curve exhibits good convergence and is almost converged with only 0.7 million k-points,
without any divergence at the Weyl point energy.
The standard deviation of the results also supports these findings, as shown in \fref{fsur-sea-dipole}(b).

%Any slight loss of symmetry in the Wannier function will exacerbate the impact of the Weyl point. 
%Although the Wannier function generally agrees well with results from density functional theory, 
%some non-negligible changes may arise for a single state at a specific k point
%such as a gapped degenerate point, unusual Berry curvature, or a non-symmetric Fermi surface.
%In \eq{dp-surf}, when using a lower k-grid, any changes to a k-point on the Fermi surface due to symmetry breaking cannot be ignored. 
%However, in \eq{dp-sea}, the sum over all occupied bands in the Brillouin zone enables a high tolerance 
%for any small symmetry losses.
%This makes \eq{dp-sea} vastly more robust than \eq{dp-surf}.

\stm{For future, we may add discussion and figures on how (badly) the fermi-surface Berry dipole converges to tracelesness.
In turn, how the value of fermi-sea integrals converges to zero when the fermi level is in the gap. }

\section{Conclusions}

In this paper, we present a Wannier interpolation scheme of the non-Abelian 
Berry curvature and orbital magnetic moment matrices for a group of bands.
This method involves grouping the bands of interest together in specific energy range,
which not only reduces computational complexity but also avoids convergence 
difficulties caused by intersections of bands within the group.
By computing the trace of interpolated quantities, the sum of the quantities of the band in 
the group are obtained.

When studying higher-order transport phenomena using ``Berry-Boltzmann equation'', 
we can use integration by parts to transfer the derivative from the distribution function 
to other quantities, thus converting the conversion from ``Fermi surface'' integration to 
``Fermi sea'' integration. The advantage of ``Fermi sea'' integration is that 
all electron states below the Fermi level contribute to the integral, 
and there is no need to evaluate the Fermi surface with dense k-sampling, 
which results in fewer k-points needed to obtain accurate results. 
In our \textit{ab initio} simulation of the Berry curvature dipole in Te, 
the ``Fermi sea'' integration demonstrates good stability and convergence 
compared to the ``Fermi surface'' integration.

However, the application of the developed method is not limited to 
improving convergence of a Berry curvature dipole. In the study of 
magnetotransport within Berry-Boltzmann formalism one gets terms where the derivatives of Berry curvature 
and orbital moment cannot be avoided. In particular, that allowed us to evaluate the recently measured \cite{Rikken19,calavalle2022gate}
electrical magnetochiral anisotropy (eMChA) in tellurium. These calculations are described in \cite{Te-eMChA-to-be-published}
%Since Wannier functions have ``gauge freedom, we use ``gauge-covariant'' derivatives 
%when interpolating the derivative matrix to protect the it. 
%The ``gauge invariant'' trace of band group are also preserved.

%%% Local Variables:
%%% mode: latex
%%% TeX-master: "pap"
%%% End:

%\input{acknowledgements}
\bibliography{bib}

\newpage
\appendix

\section{Tracelessness of Berry dipole}
\secl{traceless}
As noted before, \eq{Omega-A-grad-1} is symmetric under permutation of indices $b\leftrightarrow c$,
\beq
\Omega_{\rm A}^{ab,c}=\Omega_{\rm A}^{ac,b}\,.
\eql{Omega-symm}
\eeq
To make sense of this symmetry, let us recast the Berry curvature an a
pseudovector, $\Omega^c_{\rm A}=\epsilon_{abc}\Omega^{ab}_{\rm A}/2$.
Applying $\partial_c$ to both sides of this relation and then using
\eq{Omega-symm} gives
\beq
\bnabla_\k\cdot\bOmega_{\rm A}=0\,.
\eql{div}
\eeq
Thus, the net Berry curvature of a group of states is divergence-free
everywhere in the BZ. This is a known result for a single band, because 
Berry curvature is a curl of  Berry connection 
$\bOmega_n = i\bnabla_\k \times \ip{n}{\bnabla_\k n}$,
and a divergence of a curl is always zero. 
  This is true even at chiral touching points,
where the Berry curvature of an individual band has nonzero
divergence.  The reason is that each chiral node acts as a monopole
source of Berry curvature for one of the touching bands and as a sink
for the other; but since degenerate states must belong to the same
group A or B (recall that the two groups were assumed to be separated
in energy), the monopoles cancel each other out.  It follows from
\eq{div} that the ``Berry curvature dipole'
tensor'' \eqref{eq:dp-sea}
is always traceless. Moreover the integrand of \eq{dp-sea} is 
traceless at every k-point, therefore the resulting tensor has a zero trace 
even if integration is done on a coarse $\kk$-grid. In turn, the integrand
of \eq{dp-surf} is not traceless, therefore the tensor $\mathcal{D}_{cd}^{\rm surf}$
becomes traceless only after integration is performed with high precision. 
Naturally, in an accurate calculation  $\mathcal{D}_{cd}^{\rm surf}=\mathcal{D}_{cd}^{\rm sea}$,
however that level of precision is not always easy to achieve.

\section{Proof of chain rule}
 \secl{chain-proof}
We assume that the function $f$ is smooth, and a Taylor expansion is valid:
\beq
f(x) = f_0 + f_1x +f_2x^2+f_3 x^3 + f_4 x^4+\ldots
\eeq
Therefore $F_{ij}$ can be rewritten as 
\beq
\wt f_{ij} = f_0 + f_1 \Ham_{ij} +f_2 \Ham_{ii'} \Ham_{i'j}+f_3  \Ham_{ii'} \Ham_{i'i''}\Ham_{i''j}+
%f_4  \Ham_{ii'} \Ham_{i'i''}\Ham_{i''i'''}\Ham_{i'''j}+
\ldots
\eeq
Here we assume that $i,i',i'',\ldots,j$ all belong to the same subspace (A or B)
Which will be valid in any gauge that does not mix the subspaces.
Therefore, using the product rule we can easily take the generalized derivative
\begin{multline}
\wt f_{ij}^{:a} = f_1 \Ham^{:a}_{ij} +f_2 \left( \Ham_{ii'}^{:a} \Ham_{i'j}+ \Ham_{ii'} \Ham_{i'j}^{:a} \right)\\
    +f_3 \left(  \Ham_{ii'}^{:a} \Ham_{i'i''}\Ham_{i''j}+\Ham_{ii'} \Ham_{i'i''}^{:a}\Ham_{i''j}+\Ham_{ii'} \Ham_{i'i''}\Ham_{i''j}^{:a}\right)\\
%+f_4 \left(   \Ham_{ii'}^{:a} \Ham_{i'i''}\Ham_{i''i'''}\Ham_{i'''j}  +\Ham_{ii'} \Ham_{i'i''}^{:a}\Ham_{i''i'''}\Ham_{i'''j} + \Ham_{ii'} \Ham_{i'i''}\Ham_{i''i'''}^{:a}\Ham_{i'''j} +\Ham_{ii'} \Ham_{i'i''}\Ham_{i''i'''}\Ham_{i'''j}^{:a}   \right) 
+ \ldots
\end{multline}
Which in the Hamiltonian gauge is simplified to
\begin{multline}
    \wt f_{ij}^{:a} = \Ham^{:a}_{ij} \bigl[ f_1+ f_2(\varepsilon_i+\varepsilon_j)+f_3(\varepsilon_i^2+\varepsilon_i\varepsilon_j+\varepsilon_j^2)
%+f_4(E_i^3+E_i^2E_j+E_iE_j^2+E_j^3)   
    +\ldots  \bigr] 
\end{multline}
In case $\varepsilon_j=\varepsilon_j=\varepsilon$ it reduces to 
\begin{equation}
    \wt f_{ij}^{:a} =\Ham^{:a}_{ij} \bigl[ f_1+ 2f_2 \varepsilon+ 3f_3 \varepsilon^2+\ldots \bigr] = \Ham^{:a}_{ij} \frac{df(\varepsilon)}{d\varepsilon}
\eql{chainrule-degen}
\end{equation}
Now, if $\varepsilon_i\neq \varepsilon_j$, we can multiply and divide by $\varepsilon_i-\varepsilon_j$ and get 
\begin{multline}
\wt f_{ij}^{:a} = \Ham^{:a}_{ij} \cdot \tfrac{ f_1(\varepsilon_i-\varepsilon_j)+ f_2(\varepsilon_i^2+\varepsilon_j^2)+ f_3(\varepsilon_i^3-\varepsilon_j^3) +\ldots }{\varepsilon_i-\varepsilon_j}  = \\
= \Ham^{:a}_{ij} \cdot \tfrac{f(\varepsilon_i) - f(\varepsilon_j)}{\varepsilon_i-\varepsilon_j} 
\eql{chainrule-nondegen}
\end{multline}
Note, that in the limit $\varepsilon_i\to \varepsilon_j$ \eqref{eq:chainrule-nondegen} reduces to \eqref{eq:chainrule-degen}

%%% Local Variables:
%%% mode: latex
%%% TeX-master: "pap"
%%% End:

\section{ $\overline{B}$ inside the frozen window  \secl{Bfrozen}}
From \eqs{O-mn}{Bww} it follows that 
\beq
\overline{B}^b_{ln} = i\me{ u_l}{\Ham}{ \partial_b u_j\ww} U_{jn}
\eeq
And, if 
\beq
\Ham\ket{u_l} = \el \ket{u_l}
\eql{HuEu}
\eeq
then we may write 
\beq
\overline{B}^b_{ln} = \el \overline{A}_{ln}^{b}  \eql{B-frozen}
\eeq
However, in Wannier interpolation \eq{HuEu} is true only if the energy $\el$ is inside the frozen window. 
Otherwise, the wavefunctions $\ket{u_l}$ are not guaranteed to be eigenstates of the Hamiltonianm although the columns 
of $U$ are eigenvectors of $\Ham\ww$. Therefore, it is convenient, and computationally effective 
\stm{was it ever tested that such substitution actually improves interpolation?}
to evaluate $B^b_{ln}$ using \eq{B-frozen} when $\el$ lies in the frozen window, and use \eqs{O-mn}{Bww} otherwise.
However, such substitution becomes discontinuous at the points when a band crosses the borders of frozen window, 
which disables differentiation. 
So, we can intoroduce a smoothed version:
\begin{equation}
\wt{\overline{B}}{}^{b}_{ln}=f(\el)\el\overline{A}^b_{ln}+(1-f(\el))\overline{B}^{b}_{ln}
\end{equation}
Where $f(\varepsilon)$ is a smooth function, such that $f(\varepsilon)=1$ deep inside the frozen window, and $f(\varepsilon)=0$ far outside it.
The width of the transition region should be chosen rather narrow, but finite. Now, in order to be to use the rules 
of covariant derivative, let us write it in covariant form as
\begin{equation}
\wt{\overline{B}}{}^{b}_{ln}=f_{ll'}\Ham_{l'l''}\overline{A}^b_{l''n}+(1-f_{ll'})\overline{B}^{b}_{l'n}
\end{equation}
where  $f_{ll'}=\delta_{ll'}f(\el)$
To this equation we can directly apply the product rule \eqref{eq:gender-prod-rule}  
 and get
\xlm{Changed $V){ll'}^d$ to $\Ham_{l'l''}^{;d}$}
\begin{multline}
\wt{\overline{B}}{}^{b;d}_{ln}= 
    f_{ll'} \Ham_{l'l''}^{;d}\overline{A}^b_{l''n} +
 f^{;d}_{ll'} \Ham_{l'l''} \overline{A}^b_{l''n}-f^{;d}_{ll'}\overline{B}^{b}_{l'n} +\\
 + f_{ll'}\Ham_{l'l''}\overline{A}^{b;d}_{l''n} + \bigl(1-f_{ll'})\overline{B}^{b;d}_{l'n} 
\end{multline}
where $f_{ll'}$ is evaluated according to chainrule \eqref{eq:gender-chain-rule}.

%%% Local Variables:
%%% mode: latex
%%% TeX-master: "pap"
%%% End:

\section{Evaluation of matrices in Wanier gauge}
\secl{wannier-gauge}

%\stm{we may comment out the middle line of each subequation}

\beas{Oww-2}
\left(A\ww\right)^{a}_{ij}&\equiv&  i\ip{ u_i\ww}{\partial_a u_j\ww}  \nonumber\\ 
%   &=& \sum_\mathbf{R} e^{i\kk(\R+\bt_j-\bt_i)}\memnR{\hat{r}_a-\RR-\bt_j } = \nonumber \\
    &=&  e^{i\kk(\bt_j-\bt_i)}\sum_\mathbf{R} e^{i\kk\R}  \mathbb{A}^a_{ij}(\RR) \eql{Aww-2}  \\
\left(B\ww\right)^{a}_{ij}&\equiv&  i\me{ u_i\ww}{\Ham}{\partial_a u_j\ww} \nonumber\\
% &=& \sum_\mathbf{R} e^{i\kk(\R+\bt_j-\bt_i)}\memnR{\Ham\cdot(\hat{r}_a-\RR-\bt_j) } = \nonumber\\
 &=& e^{i\kk(\bt_j-\bt_i)}\sum_\mathbf{R} e^{i\kk\mathbf{R}}\mathbb{B}^a_{ij}(\RR)  \eql{Bww-2} \\
\Hwan^{ab}_{ij}&\equiv&  \me{\partial_a u_i\ww}{\Ham}{\partial_b u_j\ww} \nonumber\\ 
%   &=& \sum_\mathbf{R} e^{i\kk(\R+\bt_j-\bt_i)}\memnR{(\hat{r}_a-\bt_i)\cdot \Ham\cdot(\hat{r}_a-\RR-\bt_j) } = \nonumber\\
 &=& e^{i\kk(\bt_j-\bt_i)}\sum_\mathbf{R} e^{i\kk\mathbf{R}}\mathbb{C}^a_{ij}(\RR)  \eql{Hww-2} \\
\Fwan^{ab}_{ij}&\equiv&  \ip{\partial_a u_i\ww}{\partial_b u_j\ww}  \nonumber\\ 
 &=& e^{i\kk(\bt_j-\bt_i)}\sum_\mathbf{R} e^{i\kk\mathbf{R}}\mathbb{F}^{ab}_{ij}(\RR)  \eql{Fww-2} 
\eeas
\stm{is it ok to have $\Hwan$ expressed through $\mathbb{C}$ and not $\mathbb{H}$}
Where
\beas{O-R}
\mathbb{A}^a_{ij}(\RR)    & =& \memnR{\hat{r}_a-t_j^a } \\
\mathbb{B}^a_{ij}(\RR)    & =& \memnR{\Ham \cdot(\hat{r}_a-\hat{R}_a -t_j^a)}\\
\mathbb{C}^{ab}_{ij}(\RR) & =& \memnR{(\hat{r}_a-t_i^a)\cdot \Ham \cdot (\hat{r}_b-t_j^a-\hat{R}_b ) }\\
\mathbb{F}^{ab}_{ij}(\RR) & =& \memnR{(\hat{r}_a-t_i^a)\cdot (\hat{r}_b-t_j^b-\hat{R}_b ) }
\eeas
These matrix elements can be computed 
using finite-difference schemes based on the ab initio bandstructures computed on regular $\kk$-grids --- 
a procedure that is widely described in the literature \cite{marzari-prb97,wang-prb06,lopez-prb12, Ryoo-SHC}
\stm{should we expand a bit about that?}

And  the derivatives will be evaluated as 
\begin{multline}
\partial_d\left(\cal{O}\ww\right)_{ij} 
  = i e^{i\kk(\bt_j-\bt_i)}  \sum_\mathbf{R} e^{i\kk\mathbf{R}}  \mathbb{O}_{ij}(\RR) (R_d+t_j^d-t_i^d) \\
    =  i(t_j^d-t_i^d)\left(\cal{O}\ww\right)_{ij} + i e^{i\kk(\bt_j-\bt_i)}  \sum_\mathbf{R} e^{i\kk\mathbf{R}}  \mathbb{O}_{ij}(\RR) R_d
\eql{Awwad}  \\
\end{multline}

%\beas{Oww-der}
%\partial_d\left(A\ww\right)^{a}_{ij} & = &i\sum_\mathbf{R} e^{i\kk\mathbf{R}}  \mathbb{A}^a_{ij}(\RR) (R_d+t_j^d-t_i^d)\eql{Awwad}  \\
%\partial_d\left(B\ww\right)^{a}_{ij}  &= & i\sum_\mathbf{R} e^{i\kk\mathbf{R}}\mathbb{B}^a_{ij}(\RR) (R_d+t_j^d-t_i^d) \eql{Bwwad} \\
%\partial_d\Hwan^{ab}_{ij} &= &i\sum_\mathbf{R} e^{i\kk\mathbf{R}}\mathbb{C}^{ab}_{ij}(\RR) (R_d+t_j^d-t_i^d)\eql{Hwwabd} \\
%\partial_d\Hwan^{c}_{ij} &= & i\sum_\mathbf{R} e^{i\kk\mathbf{R}}\mathbb{C}^c_{ij}(\RR) (R_d+t_j^d-t_i^d)\eql{Hwwcd} \\
%\partial_d\Fwan^{ab}_{ij} & = & i\sum_\mathbf{R} e^{i\kk\mathbf{R}} \mathbb{F}^{ab}_{ij}(\RR) (R_d+t_j^d-t_i^d) \eql{Fwwabd}\\
%\partial_d\left(\Omega\ww  \right)^{c}_{ij}& = &\abc\sum_\mathbf{R} e^{i\kk\mathbf{R}}\mathbb{A}^a_{ij}(\RR) \hat{R}_b (R_d+t_j^d-t_i^d)\eql{Owwcd}
%\eeas

%%% Local Variables:
%%% mode: latex
%%% TeX-master: "pap"
%%% End:

\section{Role of Wannier Centers}
\secl{wannier-centers}

Interesting to note that quantities \eqref{eq:O-R} were defined in \cite{wang-prb06} 
and \cite{lopez-prb12} (and in Wannier90 code) with $\bt_i=0$.
As far as inclusion of $\bt$ in the phase factors correspond to a phase choice 
of the basis set, the computed physical observables should not depend on the value of $t$.
Below we will show that the total value of $\Fcov^{ab}_{mn}$ and $\Hcov^{ab}_{mn}$ 
 defined by \eqs{Fab-wanint-all}{Hab-wanint-all} does not depend on $\bt_i$. 
However, the particular value of "internal" and "external" terms will depend on $\bt$. 

In particular, setting  $\bt_i$ equal to the the Wannier charge center $\me{0j}{\hat{\rr}}{0j}$
will make  $\Fcov^{ab,\rm (ext)}_{mn}$ vanish in the "tight-binding" limit, 
where $\me{\RR i}{\hat{\rr}}{\RR'j} = \delta_{\RR\RR'}\delta_{ij}$. 
\stm{what about 
$\Hcov^{ab}_{mn}$ ? }
Also, in the ab initio calculations, the computationally heavy "external" terms 
become smaller with this choice,thus allowing to evaluate them on a coarser grid, 
to achieve the same absolute accuracy of the total Berry curvature.

 Denoting qqunatities defined by Eqs. \ref{eq:Oww-2} at $\bt_i=0$ 
 as $\zerot{\mathbb{A}}$,   $\zerot{\mathbb{B}}$,  $\zerot{\mathbb{C}}$,  $\zerot{\mathbb{F}}$ 
we find the following relations:
\beas{conventions-connection}
\mathbb{A}^a_{ij}(\RR) &=& \zerot{\mathbb{A}}^a_{ij}(\RR) - \delta_{ij}\delta_{R,0}t_j^a\\
\mathbb{B}^a_{ij}(\RR) &=& \zerot{\mathbb{B}}^a_{ij}(\RR) - \zerot{\cal H}_{ij}(\R) t_j^a\\
\mathbb{C}^{ab}_{ij}(\RR) &=& \zerot{\mathbb{C}}^{ab}_{ij}(\RR) 
    - t_i^a\zerot{\mathbb{B}}_{ij}^b(\RR) - \zerot{\mathbb{B}}_{ji}^{a*}(-\RR) t_j^b\nonumber\\
&&    + t_i^a \zerot{\cal H}_{ij}(\R) t_j^b\\
\mathbb{F}^{ab}_{ij}(\RR) &=& \zerot{\mathbb{F}}^{ab}_{ij}(\RR) 
    - t_i^a\zerot{\mathbb{A}}_{ij}^b(\RR) - t_j^b\zerot{\mathbb{A}}_{ij}^a(\RR) \nonumber\\
    &&+ \delta_{ij}\delta_{\R,0} t_i^a t_i^b 
\eeas
Defining  $\overline{t}^a_{ij}  = U^*_{j'i}t_{j'}^a U_{j'i}$ in accordance with \eq{O-mn},
we arrive at the following relationsfpr the bar quantities:
\beas{conventions-connection-hamgauge}
\overline{A}^a_{ij} &=& \zerot{\overline{A}}^a_{ij} - \overline{t}^a_{ij}\\
\overline{D}^a_{ij} &=& \zerot{\overline{D}}^a_{ij} - i\overline{t}^a_{ij}\\
\overline{B}^a_{ij} &=& \zerot{\overline{B}}^a_{ij} - \zerot{\overline{H}}_{ij'}\overline{t}^a_{j'j}\\
\overline{C}^{ab}_{ij} &=& \zerot{\overline{C}}^{ab}_{ij} - (\zerot{\overline{B}}^{a\dagger})_{ij'}\overline{t}^b_{j'j} 
  - \overline{t}^a_{ij'} \zerot{\overline{B}}^{b}_{j'j} 
  - \overline{t}^a_{ii'}  \zerot{\overline{H}}_{i'j'} \overline{t}^b_{j'j'}
\\
\overline{F}^{ab}_{ij} &=& \zerot{\overline{F}}^{ab}_{ij} - \zerot{\overline{A}}^{a}_{ij'}\overline{t}^b_{j'j} 
- \overline{t}^a_{ij'} \zerot{\overline{A}}^{b}_{j'j} - \overline{t}^a_{ij'} \overline{t}^b_{j'j}
\eeas
Note, that summation here is performed over all wannierised states $i',j'$, no matter to which set ($A$ or $B$) 
belong the states $i,j$.
Substituting these relations into \eq{Fab-wanint-all} it is straightforward to show that 
$\Fcov^{ab}_{mn}=\zerot{\Fcov}^{ab}_{mn}$. However, for $\Fcov^{ab}_{mn}$ we get 

\begin{multline}
\Hcov^{ab}_{mn}-\zerot{\Hcov}^{ab}_{mn} = t^a_{mm'}\left(H_{m'n'}A_{n'n}^b - B_{m'n}^b\right) + \conjabmn 
\end{multline}

However this differnece vanishes if we apply \eq{B-frozen}. We should note that the relation 
\eqref{eq:B-frozen} is not automatically sattisfied upon Wanier interpolation, and therefore 
it is important to enforce it by hand, and follow the procedure prescribed by \aref{Bfrozen}

Due to independence of the final result on the values of $\bt_i$,
the definition of "Wannier center" may be understood broadly --- 
not only as the Wanier charge center $\me{0j}{\hat{\rr}}{0j}$, but also, e.g., 
as the position of an atom, on which the Wannier function is located.

%%% Local Variables:
%%% mode: latex
%%% TeX-master: "pap"
%%% End:

\section{Second derivative}
\secl{second-derivative}

Using product rule again with the quantities in \sref{der-F-H}, the second covariant derivative of $\Fcov^{ab}_{mn}$ and $\Hcov^{ab}_{mn}$ are shown as following. 

\beas{Fab-wanint-der2}
\Fcov^{ab:de}_{mn}&=& \Fcov^{ab:de}_{mn,\textrm{int}} + \Fcov^{ab:de}_{mn,\textrm{ext}} \eql{Fab-wanint-der2-all}\\
\Fcov^{ab:de}_{mn,\textrm{int}}&=& \left(D_{ml'}^{a:e} D_{l'n}^{b:d} + D_{ml'}^{a} D_{l'n}^{b:de} \right) +\conjabmn \eql{Fab-wanint-der2-int}\\
\Fcov^{ab:de}_{mn,\textrm{ext}}&=& - \Bigl[\Bigl( i\Abar^{a:e}_{ml'}D^{b:d}_{l'n} + i\Abar^a_{ml'}D^{b:de}_{l'n} \nonumber\\
 & &  + iD^{a:e}_{ml'}\Abar^{b:d}_{l'n}+ iD^a_{ml'}\Abar^{b:de}_{l'n}\nonumber \\
 & &  +\Abar_{mn'}^{a:e}\Abar_{n'n}^{b:d}+\Abar_{mn'}^a\Abar_{n'n}^{b:de} \Bigr)\nonumber\\
 & &  +\conjabmn \Bigr]+ \Fbar^{ab:de}_{mn}   \eql{Fab-wanint-der-ext2}
\eeas
\beas{Hab-wanint-der2}
\Hcov^{ab:de}_{mn}&=& \Hcov^{ab:de}_{mn,\textrm{int}} + \Hcov^{ab:de}_{mn,\textrm{ext}} \eql{Hab-wanint-der2-all}\\
\Hcov^{ab:de}_{mn,\textrm{int}}&=&
            - \Bigl[ \Bigl( D_{ml'}^{a:e} {\Ham_{l'l''}} D_{l''n}^{b:d} + D_{ml'}^{a} \Ham^{,e}_{l'l''} D_{l''n}^{b:d}\nonumber \\
            & & + D_{ml'}^{a} {\Ham_{l'l''}} D_{l''n}^{b:de} + D_{ml'}^{a:e} \Ham_{l'l''}^{,d} D_{l''n}^b \Bigl) \nonumber \\
            & & +\conjabmn \Bigr] - D_{ml'}^{a} \Ham_{l'l''}^{,d:e} D_{l''n}^b
                    \eql{Hab-wanint-der2-int}\\
\Hcov^{ab:de}_{mn,\textrm{ext}}&=& \Bigl[\Bigl(
            i(\overline{B}^\dagger)^{a:e}_{ml'}D^{b:d}_{l'n} + i(\overline{B}^\dagger)^a_{ml'}D^{b:de}_{l'n} \nonumber\\
            & & + iD^{a:e}_{ml'}\overline{B}^{b:d}_{l'n} + iD^{a}_{ml'}\overline{B}^{b:de}_{l'n}  \nonumber\\
            & & - \Abar_{mm'}^{a:e} \Ham_{m'n'} \Abar_{n'n}^{b:d} - \Abar_{mm'}^a \Ham^{,e}_{m'n'} \Abar_{n'n}^{b:d} \nonumber\\
            & & - \Abar_{mm'}^a \Ham_{m'n'} \Abar_{n'n}^{b:de} - \Abar_{mm'}^{a:e} \Ham_{m'n'}^{,d} \Abar_{n'n}^{b} \Bigl) \nonumber\\
            & & + \conjabmn  \Bigr] - \Abar_{mm'}^a \Ham_{m'n'}^{,d:e} \Abar_{n'n}^{b}  + \Hbar^{ab:de}_{mn}
\eql{Hab-wanint-der2-ext}
\eeas

The derivatives $\Abar^{b:de}_{ln}$, $\overline{B}^{b:de}_{ln}$,
$\overline{H}^{b:de}_{ln}$ and $\overline{F}^{b:de}_{ln}$ are evaluated in a general way.
Take the covariant derivative agian of \eq{gender-Onn} following the pruduct rule, for
instance reads

\bea
\overline{\cal O}^{:de}_{nn'}&=&\overline{\cal O}^{,d:e}_{nn'}
-\sum_l\out D^{d:e}_{nl}{\cal O}_{ln'} -\sum_l\out D^d_{nl}{\cal O}^{:e}_{ln'}\\
& & +\sum_l\out {\cal O}^{:e}_{nl}D^d_{ln'} +\sum_l\out {\cal O}_{nl}D^{d:e}_{ln'}
\eql{gender-Onn-2nd}
\eea

And covariant matrix $D^{b:de}_{ln}$ reads

\bea
D^{b:de}_{ln}&=&\frac{1}{\varepsilon_{nl}}
\Bigl\{
\Ham^{,bd:e}_{ln} + \Bigl[\Bigl(\Ham^{,b:e}_{ll'} D^d_{l'n} + \Ham^{,b}_{ll'} D^{d:e}_{l'n}\\
& & - D^{d:e}_{ln'}\Ham^{,b}_{n'n}  - D^d_{ln'}\Ham^{,b:e}_{n'n}\Bigl) + \Bigl(b \leftrightarrow d\Bigl)\Bigl]\\
& & + \Ham^{,e}_{ll'} D^{b:d}_{ln'} - D^{b:d}_{ln'}\Ham^{,e}_{n'n}
\Bigl\}
\eql{gender-sumrule-2nd}
\eea

%%% Local Variables:
%%% mode: latex
%%% TeX-master: "pap"
%%% End:

\end{document}